\documentclass[a4paper,fleqn,usenatbib]{mnras}

\usepackage{newtxtext,newtxmath}

\usepackage[T1]{fontenc}
\usepackage{ae,aecompl}
\usepackage{units}
\usepackage{textcomp}
\usepackage{url}
\usepackage{amstext}
\usepackage{stmaryrd}
\usepackage{wasysym}
\usepackage[authoryear]{natbib}

\usepackage{graphicx}	
\usepackage{amsmath}	
\usepackage{amssymb}	
\usepackage{gensymb}
\providecommand{\tabularnewline}{\\}

\title[Deep LOFAR observations of the merging galaxy cluster CIZA J2242.8+5301]{Deep LOFAR observations of the merging galaxy cluster CIZA J2242.8+5301}

\author[D. N. Hoang et al.]{D. N. Hoang$^{1}$\thanks{E-mail: hoang@strw.leidenuniv.nl},
T. W. Shimwell$^{1}$,
A. Stroe$^{2}$,
H. Akamatsu$^{3}$,
G. Brunetti$^{4}$,\newauthor
J. M. F. Donnert$^{5,4,1}$,
H. T. Intema$^{1}$,
D. D. Mulcahy$^{6}$,
H. J. A. R\"{o}ttgering$^{1}$,\newauthor
R. J. van Weeren$^{7}$,
A. Bonafede$^{4,8}$,
M. Br\"{u}ggen$^{8}$,
R. Cassano$^{4}$,
K. T. Chy\.zy$^{9}$,\newauthor
T. En{\ss}lin$^{10,11}$, 
C. Ferrari$^{12}$,
F. de Gasperin$^{1}$,
L. Gu$^{3}$,
M. Hoeft$^{13}$,
G. K. Miley$^{1}$,\newauthor
E. Orr\'u$^{14,15}$,
R. Pizzo$^{14}$ and
G. J. White$^{16,17}$\\
$^{1}$Leiden Observatory, Leiden University, PO Box 9513, NL-2300 RA Leiden, The Netherlands\\
$^{2}$European Southern Observatory, Karl-Schwarzschild-Str. 2, 85748, Garching, Germany; ESO Fellow\\ 
$^{3}$SRON Netherlands Institute for Space Research, Sorbonnelaan 2, 3584 CA Utrecht, The Netherlands\\ 
$^{4}$INAF-Istituto di Radioastronomia, via P. Gobetti 101, I-40129 Bologna, Italy\\ 
$^{5}$School of Physics and Astronomy, University of Minnesota, Minneapolis, MN 55455, USA\\ 
$^{6}$Jodrell Bank Centre for Astrophysics, Alan Turing Building, School of Physics and Astronomy, The University of Manchester,\\
Oxford Road, Manchester, M13 9PL, UK\\ 
$^{7}$Harvard-Smithsonian Center for Astrophysics, 60 Garden Street, Cambridge, MA 02138, USA; Clay Fellow\\ 
$^{8}$University of Hamburg, Hamburger Sternwarte, Gojenbergsweg 112, 21029 Hamburg, Germany\\ 
$^{9}$Astronomical Observatory, Jagiellonian University, ul. Orla 171, 30-244 Krak\'ow, Poland\\
$^{10}$Max Planck Institute for Astrophysics, Karl-Schwarzschild-Str. 1, 85748 Garching, Germany\\
$^{11}$Ludwig-Maximilians-Universit\"{a}t M\"{u}nchen, Geschwister-Scholl-Platz 1, 80539, M\"{u}nchen, Germany\\
$^{12}$Universit\'e C\^ote d'Azur, Observatoire de la C\^ote d'Azur, CNRS, Laboratoire Lagrange, France\\
$^{13}$Th\"uringer Landessternwarte, Sternwarte 5, 07778 Tautenburg, Germany\\
$^{14}$Netherlands Institute for Radio Astronomy (ASTRON), P.O. Box 2, 7990 AA Dwingeloo, The Netherlands\\
$^{15}$Department of Astrophysics, Institute for Mathematics, Astrophysics and Particle Physics (IMAPP),\\
Radboud University Nijmegen,P.O. Box 9010, 6500 GL Nijmegen, The Netherlands\\
$^{16}$Department of Physical Sciences, The Open University, Walton Hall, Milton Keynes MK7 6AA, England\\
$^{17}$RAL Space, The Rutherford Appleton Laboratory, Chilton, Didcot, Oxfordshire OX11 0NL, England\\
}
\date{Accepted 2017. Received 2017; in original form ZZZ}

\pubyear{2017}

\begin{document}
\label{firstpage}
\pagerange{\pageref{firstpage}--\pageref{lastpage}}
\maketitle

\begin{abstract}
Previous studies have shown that
CIZA J2242.8+5301 (the 'Sausage' cluster, $z=0.192$) is a massive
merging galaxy cluster that hosts a radio halo and multiple relics.
In this paper we present deep, high fidelity, low-frequency images
made with the LOw-Frequency Array (LOFAR) between 115.5 and 179 MHz.
These images, with a noise of $140\,\mbox{\ensuremath{\mu}Jy/beam}$
and a resolution of $\theta_{\text{beam}}=7.3\arcsec\times5.3\arcsec$,
are an order of magnitude more sensitive and five times higher resolution
than previous low-frequency images of this cluster. We combined the
LOFAR data with the existing GMRT (153, 323, 608 MHz) and WSRT (1.2,
1.4, 1.7, 2.3 GHz) data to study the spectral properties of the radio
emission from the cluster. Assuming diffusive shock acceleration (DSA),
we found Mach numbers of $\mathcal{M}_{n}=2.7{}_{-0.3}^{+0.6}$ and
$\mathcal{M}_{s}=1.9_{-0.2}^{+0.3}$ for the northern and southern
shocks. The derived Mach number for the northern shock requires an
acceleration efficiency of several percent to accelerate electrons
from the thermal pool, which is challenging for DSA. Using the radio
data, we characterised the eastern relic as a shock wave propagating
outwards with a Mach number of $\mathcal{M}_{e}=2.4_{-0.3}^{+0.5}$,
which is in agreement with $\mathcal{M}_{e}^{X}=2.5{}_{-0.2}^{+0.6}$
that we derived from Suzaku data. The eastern shock is likely to be
associated with the major cluster merger. The radio halo was measured
with a flux of  $346\pm64\,\text{mJy}$ at $145\,\text{MHz}$. Across
the halo, we observed a spectral index that remains approximately constant ($\alpha^{\text{145 MHz-2.3 GHz}}_{\text{across \(\sim\)1 Mpc}^2}=-1.01\pm0.10$)
after the steepening in the post-shock region of the northern relic.
This suggests a generation of post-shock turbulence that re-energies
aged electrons.
\end{abstract}




\section{Introduction}
\label{sec:Introduction}

Diffuse Mpc-scale synchrotron emission has been observed in a number
of galaxy clusters, revealing the prevalence of non-thermal components
in the intra-cluster medium (ICM). This diffuse radio emission is
not obviously associated with compact radio sources (e.g. galaxies)
and is classified  as two groups: radio halos and relics (e.g. see
a review by \citealt{Luigina2012}). Radio halos often have a regular
shape, approximately follow the distribution of the X-ray emission,
and are apparently unpolarised. Radio relics often have an elongated
morphology, are found in the cluster outskirts, and are strongly polarised
at high frequencies. In the framework of hierarchical structure formation,
galaxy clusters grow through a sequence of mergers of smaller objects
(galaxies and sub-clusters). During merging events most of the gravitational
energy is converted into thermal energy of the ICM, but a small fraction
of it goes into non-thermal energy that includes relativistic electrons
and large-scale magnetic fields. Energetic merging events leave observable
imprints in the ICM such as giant shock waves, turbulence, and bulk
motions whose signatures are observable with radio and X-ray telescopes
(e.g. \citealt{Brunetti2014,Bruggen2012}).

The (re-)acceleration mechanisms of relativistic electrons are still
disputed for both radio halos and radio relics. There are two prominent
models that have been proposed to explain the mechanisms in radio
halos. (\textit{i}) The re-acceleration model asserts that electrons
are accelerated by turbulence that is introduced by cluster mergers
(e.g. \citealt{Brunetti2001,Petrosian2001a}). (\textit{ii}) The \textit{secondary}
acceleration model proposes that the relativistic electrons/positrons
are the \textit{secondary} products of hadronic collisions between
relativistic protons and thermal ions present in the ICM (e.g. \citealt{Dennison1980a,Blasi1999,Dolag2000}).
The former model is thought to generate radio emission that is observable
for approximately 1 Gyr after major merging events (\citealt{brunetti2009,Miniati2015}).
In the latter model the radio emission is sustained over the lifetime
of a cluster due to the long lifetime of relativistic protons in the
ICM leading to the continuous injection of secondary particles. The
\textit{secondary} model also predicts the existence of $\gamma-$rays
as one of the products of the decay chain associated with hadronic
collisions. But despite numerous studies with the \textit{Fermi Gamma-ray
Space Telescope} (e.g. \citealt{Jeltema2011,Brunetti2012,Zandanel2014,Ackermann2016b}),
no firm detection of the $\gamma-$rays from the ICM has been challenging
this scenario. Still secondary electrons may contribute to the observed
emission, for instance a hybrid model where turbulence re-accelerates
both primary particles and their secondaries has also been proposed
to explain radio halos (\citealt{Brunetti2004,Brunetti2011b,Pinzke2016});
in this case the expected $\gamma-$ray  emission is weaker than that
expected in purely secondary models.

Radio relics are generally thought to trace shock waves in the cluster
outskirts that are propagating away from the cluster after a merging
event (e.g. \citealt{EnBlin1998,Roettiger1999a}). It is also thought
that some radio relics might be generated by shocks associated with
in-falling matter from cosmic filaments (e.g. \citealt{EnBlin1998,EnBlin2001,Brown2011}).
Particle acceleration at shocks can be described by the diffusive
shock acceleration (DSA) model (e.g. \citealt{Bell1978,Drury1983,Blandford1987}).
However shocks in galaxy clusters are weak (Mach $\lesssim5$) and
in some cases the plausibility of the acceleration of thermal particles
in the ICM by DSA is challenged by the observed spectra of radio relics
and by the efficiencies that would be required to explain observations
(e.g. see \citealt{Brunetti2014} for review, \citealt{Akamatsu2015,Vazza2015,VanWeeren2016b,Botteon2016}).
However, these problems can be mitigated if the shock re-accelerates
fossil electrons that have already been accelerated prior to the merging
event (e.g. \citealt{Markevitch2005,Kang2011a,Kang2012}). Obvious
candidate sources of fossil electrons are radio galaxies on the outskirts
of the relic cluster. Observationally, this re-acceleration mechanism
was proposed to explain the radio emission in a few clusters such
as Abell 3411-3412 (\citealt{weeren2013,VanWeeren2017}), PLCKG287.0
+32.9 (\citealt{Bonafede2014}) and the Bullet cluster 1E 0657\textminus 55.8
(\citealt{shimwell2015}).

\subsection*{The galaxy cluster CIZA J2242.8+5301}

CIZA J2242.8+5301 (hereafter CIZA2242, $z=0.192$) is a massive galaxy
cluster that hosts an excellent example of large-scale particle acceleration.
CIZA2242 was originally discovered in the ROSAT All-Sky Survey and
was identified as a galaxy cluster undergoing a major merger event
by \citet{Kocevski2007}. The cluster has since been characterised
across a broad range of electromagnetic wavelengths including X-ray,
optical and radio, and its properties have been interpreted with the
help of simulations. 

\textit{XMM-Newton} X-ray observations (\citealt{Ogrean2013a}) confirmed
the merging state of the cluster and characterised its disturbed morphology
and elongation in the north-south direction. Suzaku observations (\citealt{Akamatsu2013,Akamatsu2015})
detected an ICM temperature jump, indicating the presence of merger
shocks in the north and south of the cluster. The Mach numbers of
these shocks were estimated as $\mathcal{M}_{n}=2.7_{-0.4}^{+0.7}$
and $\mathcal{M}_{s}=1.7_{-0.3}^{+0.4}$, respectively. Chandra observations
(\citealt{Ogrean2014}) revealed additional discontinuities in the
X-ray surface brightness in multiple locations in the cluster outskirts
(see Fig. 8 in \citealt{Ogrean2014}). In the optical band, a comprehensive
redshift analysis to study the geometry and dynamics of the merging
cluster \citet{Dawson2015} found that CIZA2242 consists of two sub-clusters
that are at similar redshift but have virtually no difference in the
line-of-sight velocity ($69\pm190\,\mbox{km\,\ensuremath{s^{-1}}}$)
and are separated by a projected distance of $1.3_{-0.10}^{+0.13}\,\mbox{Mpc}$.

Radio observations with the GMRT (at 608 MHz) and WSRT (at 1.2, 1.4,
1.7, and 2.3 GHz) reported two opposite radio relics located at the
outskirts ($1.5\,\mbox{Mpc}$ from the cluster centre, \citealt{VanWeeren2010a}).
The northern relic has an arc-like morphology, a size of $2\,\mbox{Mpc}\times55\,\mbox{kpc}$,
spectral index gradients from $-0.6$ to $-2.0$ across the width
of the relic and a high degree of polarisation ($50-60\%$, VLA $4.9\,\text{GHz}$
data). The relics have been interpreted as tracing shock waves propagating
outward after a major cluster merger. The injection spectral index
of $-0.6\pm0.05$ of the northern relic, that was calculated from
the radio observations, corresponds to a Mach number of $4.6_{-0.9}^{+1.3}$
and is higher than the values derived from X-ray studies (e.g. $\mathcal{M}_{n}^{X}=2.54_{-0.43}^{+0.64}$
in \citealt{Ogrean2014}). The magnetic field strength was estimated
to be within $5-7\,\mbox{\ensuremath{\mu}G}$ to satisfy the conditions
of the spectral ageing, the relic geometry and the ICM temperature.
Faint emission connecting the two relics was detected in the WSRT
1.4 GHz map and was interpreted as a radio halo by \citet{VanWeeren2010a}
but was not characterised in detail. \citet{Stroe2013a} performed
further studies of CIZA2242 using GMRT 153 and 323 MHz data, in combination
with the existing data. Integrated spectra for the relics were reported,
and by using standard DSA/re-acceleration theory, \citet{Stroe2013a}
estimated Mach numbers of $\mathcal{M}_{n}=4.58\pm1.09$ for the northern
radio relic (from the injection index of $-0.6\pm0.05$ which they
obtained from colour-colour plots) and $\mathcal{M}_{s}=2.81\pm0.19$
for the southern radio relic (derived from the integrated spectral
index of $-1.29\pm0.04$ using DSA model). \citet{Stroe2013a} found
variations in the radio surface brightness on scales of 100 kpc along
the length of the northern relic and linked them with the variations
in ICM density and temperature (\citealt{Hoeft2011}). Additionally
\citet{Stroe2013a} reported relics on the eastern side of the cluster
and characterised 5 tailed radio galaxies spread throughout the ICM.

Despite CIZA2242 being an exceptionally well-studied cluster, several
questions remain unanswered, such as (\textit{i}) the discrepancy
between the radio and X-ray derived Mach numbers for the northern
and southern relics; (\textit{ii}) the connection between the radio
halo and the northern and southern relics; (\textit{iii}) the spectral
properties of the radio halo, southern and eastern relics; (\textit{iv})
the nature of the eastern relics. In this paper we present LOFAR (\citealt{VanHaarlem2013})
observations of CIZA2242 using the High Band Antenna (HBA). With its
excellent surface brightness sensitivity coupled with high resolution,
LOFAR is well-suited to study objects that host both compact and very
diffuse emission, such as CIZA2242. The high density of core stations
is essential for the detection of diffuse emission from CIZA2242 which
has emission on scales of up to $17'$. In this paper we offer new
insights into the above questions by exploiting our high spatial resolution,
deep LOFAR data in combination with the published GMRT, WSRT, Chandra
and Suzaku data (\citealt{VanWeeren2010a,Stroe2013a,Ogrean2014,Akamatsu2015}).

Hereafter we assume a flat cosmology with $\Omega_{M}=0.3$, $\Omega_{\Lambda}=0.7$,
and $H_{0}=70$ km s$^{-1}$ Mpc$^{-1}$. In this
cosmology, an angular distance of $1'$ corresponds to a physical
size of 192 kpc at $z=0.192$. In this paper, we use the convention
of $S\propto\nu^{\alpha}$ for radio synchrotron spectrum, where $S$
is the flux density at frequency $\nu$ and $\alpha$ is the spectral
index.

\section{Observations and data reduction}
\label{sec:Obs}

\subsection{LOFAR HBA data}

CIZA2242 was observed with LOFAR during the day for 9.6 hours (8:10
AM to 17:50 PM) on February 21, 2015. The frequency coverage for the
target observation was between 115.5 MHz and 179.0 MHz. The calibrator
source 3C 196 was observed for 10 minutes after the target observation.
Both observations used 14 remote and 46 (split) core stations (see
\citealt{VanHaarlem2013} for a description of the stations), the
baseline length range is from 42 m to 120 km. A summary of the observations
is given in Table \ref{tab:SS_obs_table}.

\begin{table}
\caption{LOFAR HBA observation parameters \label{tab:SS_obs_table}}

\begin{tabular}{lc}
 \hline 
Observation IDs & L260393 (CIZA2242), L260397 (3C 196)\tabularnewline
Pointing centres & 22:42:53.00, +53.01.05.01 (CIZA2242),\tabularnewline
 & 08:13:36.07, +48.13.02.58 (3C 196)\tabularnewline
Integration time & 1 s\tabularnewline
Observation date & February 21, 2015\tabularnewline
Total on-source time & 9.6 hr (CIZA2242),\tabularnewline
 & 10 min (3C 196)\tabularnewline
Correlations & XX, XY, YX, YY\tabularnewline
Frequency range  & 115.5-179.0 MHz (CIZA2242)\tabularnewline
 & 109.7-189.9 MHz (3C 196)\tabularnewline
Total bandwidth & 63.5 MHz (CIZA2242,\tabularnewline
 &  usable 56.6 MHz)\tabularnewline
Total number of & \tabularnewline
 sub-band (SB) & 325 (CIZA2242, usable 290 SBs)\tabularnewline
Bandwidth per SB & 195.3125 kHz\tabularnewline
Channels per SB & 64\tabularnewline
Number of stations & 60 (46 split core + 14 remote)\tabularnewline
 \hline 
\end{tabular}
\end{table}

To create high spatial resolution, sensitive images with good fidelity
direction-independent calibration and direction-dependent calibration
were performed. The direction-independent calibration of the target
field aims to (\textit{i}) remove the contamination caused by radio
frequency interference (RFI) and the bright sources (e.g. Cassiopeia
A, Cygnus A) located in the side lobes, (\textit{ii}) to correct the
clock offset between stations and (\textit{iii}) to calibrate the
XX-YY phase of the antennas. For the direction dependent part, we
used the recently developed facet calibration scheme that is described
in \citet{VanWeeren2016a}.

Throughout the data reduction process, we used BLACKBOARD SELFCAL
(BBS, \citealt{Pandey2009}) for calibrating data, LOFAR Default PreProcessing
Pipeline (DPPP) for editing data (flag, average, concatenate), and
w-Stacking Clean (WSClean, \citealt{Offringa2014}), Common Astronomy
Software Applications (CASA, \citealt{Mullin2007}) and AW{\scriptsize{}IMAGER}
(\citealt{Tasse2013}) for imaging.

\subsubsection{Direction-independent calibration \label{subsec:Non-directional-calibration}}
\begin{itemize}
\item Removal of RFI
\end{itemize}
The data of CIZA2242 and 3C 196 were flagged to remove RFI contamination
with the automatic flagger AOFLAGGER (\citealt{Offringa2012c}). The
auto-correlation and the noisy channels at the edge of each subband
(first and last two channels) were also removed with DPPP by the Radio
Observatory\footnote{\url{http://www.lofar.org}}. The edge channels
were removed to avoid calibration difficulties caused by the steep
curved bandpass at the edge of subbands.
\begin{itemize}
\item Removal of of distant contaminating sources
\end{itemize}
As with other low-frequency observations, the data were contaminated
by emission from strong radio sources dozens of degrees away from
the target. This contamination is predominately from  several A-team
sources: Cassiopeia A (CasA), Cygnus A (CygA), Taurus A (TauA), Hercules
A (HerA), Virgo A (VirA), and Jupiter. To remove this contamination,
we applied two different techniques depending on the angular separation
of the contaminating source and CIZA2242. Our efforts focused on the
four high-elevation sources: CasA (12.8 kJy at 152 MHz), CygA (10.5
kJy at 152 MHz), TauA (1.43 kJy at 152 MHz), and HerA (0.835 kJy at
74 MHz) which are approximately $8\degree$, $30\degree$, $79\degree$,
and $85\degree$ away from CIZA2242 location, respectively (\citealt{BaarsJ1977,Gizani2005}).
The closest source, CasA, was subtracted from the CIZA2242 data using
'demixing', a technique developed by \citet{VanderTol2007}, whereas
the other A-team sources were removed based on the amplitude of their
simulated visibilities. The former technique solves for direction-dependent
gain solutions towards CasA using an input sky model, and subtracts
the contribution of CasA from the data using these gain solutions
and the input sky model. The sky model we used for CasA was from a
high-resolution ($\sim10\arcsec$) image and contains more than $16,000$
components with an integrated flux of 30.77 kJy (at 69 MHz, R. van
Weeren, priv. comm.). The latter technique simulates visibilities
of the A-team sources (CygA, TauA, and HerA) by performing inverse
Fourier transforms of their sky models with the station beam applied
in BBS and then flags the target data if the simulated visibility
amplitudes are larger than a chosen threshold of 5 Jy.
\begin{itemize}
\item Amplitude calibration, initial clock-offset and XX-YY phase-offset
corrections
\end{itemize}
Following the procedure that is described in \citet{VanWeeren2010a},
we assumed the flux scale, clock offset and XX-YY phase offset are
direction and time independent and can be corrected in the target
field if they are derived from a calibrator observation. In this study,
3C 196 was used as a calibrator. First, the XX and YY complex gains
were solved for each antenna every 4 s and 1.5259 kHz using a sky
model of 3C 196 (V. N. Pandey, priv. comm.). The 3C 196 sky model
contains 4 compact Gaussians with a total flux of $83.1\,\text{Jy}$,
which is consistent with the \citet{Scaife2012} flux scale. In this
calibration, the Rotation Angle $\beta$ was derived to account for
the differential Faraday Rotation effects from the parallel hand amplitudes.
The LOFAR station beam was also used during the solve step to separate
the beam effects from the complex gain solutions.

For LOFAR, while the core stations use a single clock, the remote
stations have separate ones. The clocks are synchronised, but there
are still small offsets. These offsets are up to hundreds of nano-seconds.
We applied a clock-TEC separation technique to estimate the clock
offsets (see \citealt{VanWeeren2016a} for details). The XX-YY phase
offsets for each station were calculated by taking the difference
of the medians of the XX and YY phase gain solutions taken over the
whole 10-minutes observation of 3C 196.

Finally the XX-YY phase offset, the initial clock offset, and the
amplitude gains were transferred to the target data. Since the calibrator,
3C 196, is $\sim74\degree$ away from the target field, it has different
ionospheric conditions and we did not transfer the TEC solutions to
the target.
\begin{itemize}
\item Initial phase calibration and the subtraction of all sources in the
target field 
\end{itemize}
The target data sets of single subbands were concatenated to blocks
of 2-MHz bandwidth to increase S/N ratio in the calibration steps.
The blocks were phase calibrated against a wide-field sky model which
was extracted from a GMRT 153 MHz image (radius of $\sim2\degree$
and at $\sim25\arcsec$ resolution, \citealt{Stroe2013a}). Phase
solutions for each 2-MHz block were obtained every 8 s, which is fast
enough to trace typical ionospheric changes. Note that as we already
had a good model of the target field, to reduce processing time we
did not perform self-calibration of the field  as has been done in
other studies that also use the facet calibration scheme (e.g. \citealt{VanWeeren2016b}).

After phase calibration, and to prepare for facet calibration, we
subtracted all sources from the field. To do this, we made medium
resolution ($\sim30\arcsec$) images of the CIZA2242 field for each
$2\,\text{MHz}$ block with WSClean (Briggs weighting, $\textrm{robust}=0$).
The size of these images is set to $10\degree\times10\degree$ so
that it covers the main LOFAR beam. The CLEAN components, together
with the direction independent gain solutions in the previous step,
were used to subtract sources from the data. Afterwards, to better
subtract low-surface brightness emission and remove sources further
than $10\degree$ away from the location of CIZA2242, we followed
the same steps as above. But the data, which were already source subtracted,
were imaged at lower resolution ($2\arcmin$) over a wide-field area
($20\degree\times20\degree$) that encompassed the second sidelobe
of the LOFAR beam. The low-resolution sky models were subtracted from
the medium-resolution subtracted data using the direction-independent
gain solutions. The target data sets, which we hereafter refer to
as ``blank'' field datasets, now contain just noise and residuals
from the imperfect source subtraction.

\subsubsection{Direction-dependent calibration \label{subsec:Directional-calibration}}

In principle, we could directly calibrate the antenna gains and correct
for the ionospheric distortion in the direction of CIZA2242 by calibrating
off a nearby bright source. However, the imperfections in the source
subtraction that used direction-independent calibration solutions
result in non-negligible residuals in the ``blank'' field images,
especially in regions around bright sources. For this reason, we exploited
facet calibration (\citealt{VanWeeren2016a}) to progressively improve
the source subtraction in the ``blank'' data sets, and consequentially,
gradually reduce the noise in the ``blank'' field datasets as the
subtraction improves. Below we briefly outline the direction dependent
calibration procedure.

The CIZA2242 field was divided into 15 facets covering an area of
$\sim3\degree$ in radius. Each facet has its own calibrator consisting
of one or more sources that have a total apparent flux in excess of
0.5 Jy (without primary beam correction). The number of facets here
is close to that used in another cluster study by \citet{Shimwell2016a}
(13 facets), but far less than that in \citet{Williams2016a} (33
facets) and \citet{VanWeeren2016b} (70 facets). In this study we
used few facets to reduce the computational time and because we only
require high quality images of the cluster region which has radius
of $8\arcmin$, whereas \citet{Williams2016a} targeted a $19\:\textrm{deg}^{2}$
wide-field image. 

The procedure to calibrate each facet was as follows: Firstly, in
the direction of each facet calibrator we performed a self-calibration
loop to determine a single TEC and phase solution every $8-16\,\text{s}$
per station per $8\,\text{MHz}$ bandwidth, and a single gain solution
every 4-16 mins per station per $2\,\text{MHz}$ bandwidth.  Secondly,
Stokes I images of the entire facet region to which the direction-dependent
calibration solutions were applied were made using WSClean.  These
full-bandwidth (56.6 MHz) images typically had a noise level of $\sim150\,\text{\ensuremath{\mu}Jy/beam}$
and the CLEAN components derived from the imaging form a significantly
improved frequency dependent sky model for the region (in comparison
to the direction independent sky model). Thirdly, the facet sky models
were subtracted from the individual 2-MHz bandwidth data sets using
the gains and TEC solutions in the direction of the facet calibrator
that were derived during the self-calibration loop. This subtraction
was significantly improved over the direction independent subtraction.
This procedure was repeated to successfully calibrate and accurately
subtract the sources in 11 facets, including the cluster facet, which
was done last. Four of the facets failed as their facet centre is
either far away ($2.0\degree-2.7\degree$) from the  pointing centre
 or they had low flux calibrators  which prevented us from obtaining
stable calibration solutions. These failed facets had very little
effect on the quality of the final cluster image as the subtraction
of these facet sources using the low and medium resolution sky models
with the direction-independent calibration solutions was almost sufficient
to remove the artefacts across the cluster region.

\subsection{GMRT, WSRT radio, Suzaku and Chandra X-ray data}

In this paper, we used the GMRT 153, 323, 608 MHz and WSRT 1.2, 1.4,
1.7, 2.3 GHz data sets that were originally published by \citet{VanWeeren2010a}
and \citet{Stroe2013a}. For details on the data reduction procedure,
see \citet{Stroe2013a}. To study the X-ray emission from CIZA2242
we used observations from the Suzaku and Chandra X-ray satellites.
We refer to \citet{Akamatsu2015} and \citet{Ogrean2014} for the
data reduction procedure.

\subsection{Imaging and flux scale of radio intensity images \label{subsec:Imaging}}

To make the final total intensity image of CIZA2242, we ran the CLEAN
task in CASA on the full-bandwidth (56.6 MHz) data that was calibrated
in the direction of the target. The imaging was done with multiscale-multifrequency
(MS-MFS) CLEAN, multiple Taylor terms ($\text{nterms}=2$) and W-projection
options to take into account of the complex structure, the frequency
dependence of the wide-bandwidth data sets and the non-coplanar effects
(e.g. see \citealt{Cornwell2005,Cornwell2008,Rau2011}). The multi-scale
sizes used were 0, 3, 7, 25, 60, and 150 times the pixel size, which
is approximately a fifth of the synthesised beam; the zero scale is
for modelling point sources. The multi-scale CLEAN in CASA has been
tested and shown to recover low-level diffuse emission properly, significantly
minimise the clean ``bowl'',  recover flux closer to that of single-dish
observations and leave more uniform residuals than classical single-scale
CLEAN (\citealt{Rich2008}). Several images were made using Briggs
weighting with different robust parameters to enhance diffuse emission
at different scales (see Table \ref{tab:Imaging_Parameters}). During
imaging we also applied an inner uv cut of $0.2\:k\lambda$ to filter
out the (possible) emission on scales larger than $17\arcmin$ ($\sim3.2\,\textrm{Mpc}$),
which is approximately the physical size of the cluster. The final
image was corrected for the primary beam attenuation (less than $0.5\%$
at the cluster outskirts) by dividing out the real average beam model\footnote{the square root of the AW{\scriptsize{}IMAGER} .avgpb map.}
that was produced using AW{\scriptsize{}IMAGER} (\citealt{Tasse2013}). 

The amplitude calibration was performed using the primary calibrator
3C 196 (see Subsec. \ref{subsec:Non-directional-calibration}). To
check our LOFAR flux scale, we compared the integrated fluxes of the
diffuse emission of the northern relic and two bright point-like sources
(source 1, $\sim1\,\text{Jy}$, at RA=22:41:33, Dec=+53.11.06; source
2, $\sim0.1\,\text{Jy}$, at RA=22:432:37, Dec=+53.09.16) in our LOFAR
image with the values that are predicted from spectral fitting of
the GMRT 153, 323, 608 MHz and WSRT 1.2, 1.4, 1.7, 2.3 GHz data (\citealt{Stroe2013a}).
For this comparison, we used identical imaging parameters for the
LOFAR, GMRT and WSRT data sets (see the parameters for the $16\arcsec\times18\arcsec$
images in Table \ref{tab:Imaging_Parameters}). The predicted fluxes
were found to be $S_{n}=1593\pm611\,\text{mJy}$, $S_{1}=1081\pm124\,\text{mJy}$
and $S_{2}=119\pm3\,\text{mJy}$ for the northern relic, source 1
and 2, respectively. The values that were measured within  $3\sigma_{\text{noise}}$
regions  of our LOFAR image were $S_{n}=1637\pm37\,\text{mJy}$, $S_{1}=1036\pm1\,\text{mJy}$
and $S_{2}=92\pm1\,\text{mJy}$ and are in good agreement with the
spectral fitting predicted values. This LOFAR flux for the northern
relic was only $3\%$ higher than the predicted value, and the fluxes
for source 1 and 2 were $4\%$ and $22\%$ lower than the predicted
values.  Despite of this agreement of the LOFAR, GMRT and WSRT fluxes,
throughout this paper, unless otherwise stated, we used a flux scale error of $10\%$ for all
LOFAR, GMRT, WSRT images when estimating the spectra of diffuse emission.
Similar values have been widely used in literature (e.g. \citealt{Shimwell2016a,VanWeeren2016b}).

\subsection{Spectral index maps \label{subsec:Spectral_Index_maps}}

\begin{table*}
\caption{Imaging parameters \label{tab:Imaging_Parameters}}

\begin{tabular}{cccccc}
 \hline 
Resolution & $7.3\arcsec\times5.3\arcsec$ & $6.5\arcsec\times6.5\arcsec^{a}$ & $12\arcsec\times12\arcsec^{a}$ & $16\arcsec\times18\arcsec^{a}$ & $35\arcsec\times35\arcsec^{a}$\tabularnewline
(Fig.) & (\ref{fig:SS_Hres}) & (\ref{fig:SPX_HLres}$^{b}$) & (\ref{fig:SS_Northern_region_12arcsec}) & (\ref{fig:SPX_RS_R1_R2}$^{b}$) & (\ref{fig:SS_Lres_Xray}, \ref{fig:Halo_Spx_profile}$^{b}$)\tabularnewline
 \hline 
Mode & MFS & MFS & MFS & MFS & MFS\tabularnewline
Weighting & Briggs & Uniform & Briggs & Uniform & Briggs\tabularnewline
Robust & $-0.25$ & N/A & $0.25$ & N/A & $0.5$\tabularnewline
uv-range ($\text{k\ensuremath{\lambda}}$) & $\geqslant0.2$ & $0.2-50^{c}$ & $\geqslant0.2$ & $0.2-50^{c}$ & $0.2-50^{c}$\tabularnewline
Multi-scales & $[0,3,7,60,150]$ & $[0,3,7,60,150]$ & $[0,3,7,60,150]$ & $[0,3,7,60,150]$ & $[0,3,7,60,150]$\tabularnewline
Grid mode & wide-field & wide-field & wide-field & wide-field & wide-field\tabularnewline
W-projection & $128^{d}$ & $384^{d}$, $128^{e}$  & $128^{d}$ & $384^{d}$, $128^{e}$, $256^{f}$ & $384^{d}$, $128^{e}$, $256^{f}$\tabularnewline
N-terms & $2^{d}$ & $2^{d}$, $1^{e}$ & $2^{d}$ & $2^{d}$, $1^{e,f}$ & $2^{d}$, $1^{e,f}$\tabularnewline
Image RMS & $140^{d}$ & $200^{d}$,  & $210^{d}$ & $312^{d}$, & $430^{d}$, \tabularnewline
($\text{\ensuremath{\mu}Jy/beam}$) &  & $37^{e_{3}}$ &  &  $1358^{e_{1}}$, $414^{e_{2}}$, $64^{e_{3}}$, & $2000^{e_{1}}$, $495^{e_{2}}$, $177^{e_{3}}$,\tabularnewline
 &  &  &  & $70^{f_{1}}$, $31^{f_{2}}$, $38^{f_{3}}$, $43^{f_{4}}$ & $99^{f_{1}}$, $71^{f_{2}}$, $73^{f_{3}}$, $70^{f_{4}}$\tabularnewline
 \hline 
\end{tabular}\\
$^{a}$: smoothed, $^{b}$: spectral index map, $^{c}$: $\text{uv}_{\text{max}}=50\,\text{k\ensuremath{\lambda}}$
only used for LOFAR data, $^{d}$: LOFAR, $^{e}$: GMRT ($^{e_{1}}$,
$^{e_{2}}$ and $^{e_{3}}$ are for 153, 323 and 608 MHz, respectively),
$^{f}$: WSRT ($^{f_{1}}$, $^{f_{2}}$, $^{f_{3}}$ and $^{f_{4}}$
are for 1.2, 1.4, 1.7 and 2.3 GHz, respectively)
\end{table*}

Our high-fidelity LOFAR images have allowed us to map the spectral
index distribution with improved resolution. In previous works (\citealt{VanWeeren2010a,Stroe2013a}),
CIZA2242 was studied with the GMRT and WSRT at seven frequencies from
153 MHz to 2.3 GHz. Our LOFAR 145 MHz data was combined with these
published data sets to study spectral characteristics of the cluster.
However, these observations were performed with different interferometers
each of which has a different uv-coverage, and this results in a bias
in the detectable emission and the spectra. To minimise the difference,
we (re-)imaged all data sets with the same weighting scheme of visibilities
and selected only data with a common inner uv-cut of $0.2\,\text{k\ensuremath{\lambda}}$.
To make the spectral index maps all images were made using MS-MFS
CLEAN ($\text{multiscale}=[0,3,7,25,60,150]\times\text{pixel sizes}$
and $\text{nterms}=1\,\text{and}\,2$ for GMRT/WSRT and LOFAR images,
respectively). Only those pixels with values $\geqslant3\sigma_{\text{noise}}$
in each of the individual images were used for the spectral index
calculation.  We note that this $\geqslant3\sigma_{\text{noise}}$
cut-off introduces a selection bias for steep spectrum sources. For
example, the sources that were observed with LOFAR at $\geqslant3\sigma_{\text{noise}}$
but were not detected ($<3\sigma_{\text{noise}}$) with the GMRT/WSRT
observations were not included in the spectral index maps. To reveal
spectral properties of different spatial scales, we made spectral
index maps at $6.5\arcsec$ , $18\arcsec\times16\arcsec$ and $35\arcsec$
resolution (see Table \ref{tab:Imaging_Parameters} for a summary
of the imaging parameters).

The $6.5\arcsec$-resolution spectral index map was made with the
LOFAR 145 MHz and GMRT 608 MHz data sets. The imaging used uniform
weighting for both data sets. In addition a common uv-range was used
($0.2\,k\lambda$ to $50\,k\lambda$) and a uvtaper of $6.0\arcsec$
was applied to reduce the sidelobes and help with CLEAN convergence.
Here, the $50\,\text{k\ensuremath{\lambda}}$ is the maximum uv distance
of the GMRT data set. The native images reach resolution of $\sim6\arcsec$
($5.5\arcsec\times5.3\arcsec$ for the LOFAR 145 MHz, $5.7\arcsec\times5.4\arcsec$
for the GMRT 608 MHz), which were then convolved with a 2D Gaussian
kernel to a common resolution of $6.5\arcsec$, aligned with respect
to the LOFAR image, and regrided to a common pixelisation. To align
the images, we fitted compact sources with 2D Gaussian functions to
find their locations which were used to estimate the average displacements
between the GMRT/WSRT and LOFAR images. The GMRT/WSRT images were
then shifted along the RA and DEC axes.\textbf{} The final images
were combined to make spectral index maps according to

\begin{equation}
\alpha_{\text{pixel}}=\frac{\ln\frac{S_{1}}{S_{2}}}{\ln\frac{\nu_{1}}{\nu_{2}}},\label{eq:SPX}
\end{equation}
where $S_{1}$ and $S_{2}$ are the pixel values of the LOFAR and
GMRT maps at the frequency $\nu_{1}=145\:\textrm{MHz}$ and $\nu_{2}=608\,\text{MHz}$,
respectively. We estimated the spectral index error on each pixel,
$\Delta\alpha_{\text{pixel}}$, taking into account the image noise
$\sigma_{\text{noise}}$ and the flux scale error of $f_{\text{err}}=10\%$

\begin{equation}
\Delta\alpha_{\text{pixel}}=\frac{1}{\ln\frac{\nu_{1}}{\nu_{2}}}\sqrt{\left(\frac{\Delta S_{1}}{S_{1}}\right)^{2}+\left(\frac{\Delta S_{2}}{S_{2}}\right)^{2}},\label{eq:Spec_Index_Error}
\end{equation}
where $\Delta S_{i}=\sqrt{\left(\sigma_{\text{noise}}^{i}\right)^{2}+\left(f_{\text{err}}\times S_{i}\right)^{2}}$
are the total errors of $S_{i}$. The spectral index error for a region
that covers more than one pixel and has constant spectral indices
is calculated as follows

\begin{equation}
\Delta\alpha_{\text{region}}=\frac{\sqrt{\sum_{1}^{N_{\text{beams}}}\left(\overline{\Delta\alpha_{\text{pixels}}}\right)^{2}}}{N_{\text{beams}}},\label{eq:Avg_Spx_Error}
\end{equation}
where $\overline{\Delta\alpha_{\text{pixels}}}$ is an average of
all $\Delta\alpha_{\text{pixels}}$ in the region of area of $N_{\text{beams}}$
beam sizes.

The $18\arcsec\times16\arcsec$-resolution map was made with the LOFAR
145 MHz, GMRT 153, 323, 608 MHz, and the WSRT 1.2, 1.4, 1.7, 2.3 GHz
data sets. The imaging was done with similar settings as used for
the $6.5\arcsec$-resolution map (uniform weighting, $\text{uv}_{\text{min}}=0.2\,\mbox{k\ensuremath{\lambda}}$,
$\text{uv}_{\text{max}}=50\,\text{k\ensuremath{\lambda}}$). Here a  $\text{uvtaper}=6\arcsec$
is only applied to the LOFAR and GMRT 608 MHz data sets to improve
the CLEAN convergence. All eight images were then smoothed to a common
resolution of $18\arcsec\times16\arcsec$, aligned with respect to
the LOFAR image, and regrided. To obtain the eight-frequency spectral
index map, we fitted a power-law function  to each pixel of the eight
images using a weighted least-squares technique. The fitting was done
only on pixels that have a signal of $\geqslant3\sigma_{\text{noise}}$
in least four observations. To take into account the uncertainties
of the individual maps, the pixels are weighted by the inverse-square
of their total pixel errors ($1/\Delta S_{i}^{2}$) which includes
the individual image noise $\sigma_{\text{noise}}$ and an error of
$10\%$ in the flux scale (Eq. \ref{eq:Spec_Index_Error}).

The $35\arcsec$-resolution map was made in a similar manner to the
$18\arcsec\times16\arcsec$-resolution spectral index map ($\text{uv}_{\text{min}}=0.2\,\text{k}\lambda$,
MS-MFS, W-projection). However, instead of using uniform weighting,
Briggs weighting ($\text{robust}=0.5$) was used to increase S/N ratio
of the diffuse emission associated with CIZA2242. The outer taper
used for each image was set to obtain a spatial resolution of nearly
$30\arcsec$. The images were then convolved with a 2D Gaussian kernel
to give images with a common resolution of $35\arcsec$, aligned with
respect to the LOFAR image, and regrided to have the same pixel size.
The $35\arcsec$ spectral index and corresponding error maps between
145 MHz and 2.3 GHz were made following the procedure that was used
for the $18\arcsec\times16\arcsec$-resolution maps, except that the
minimum number of detections ($\geqslant2\sigma_{\text{noise}}$)
was limited to three, rather than four, images.

\section{Results}
\label{sec:Results}
 
\begin{figure*}
\centering{}\includegraphics[clip,width=0.80\textwidth]{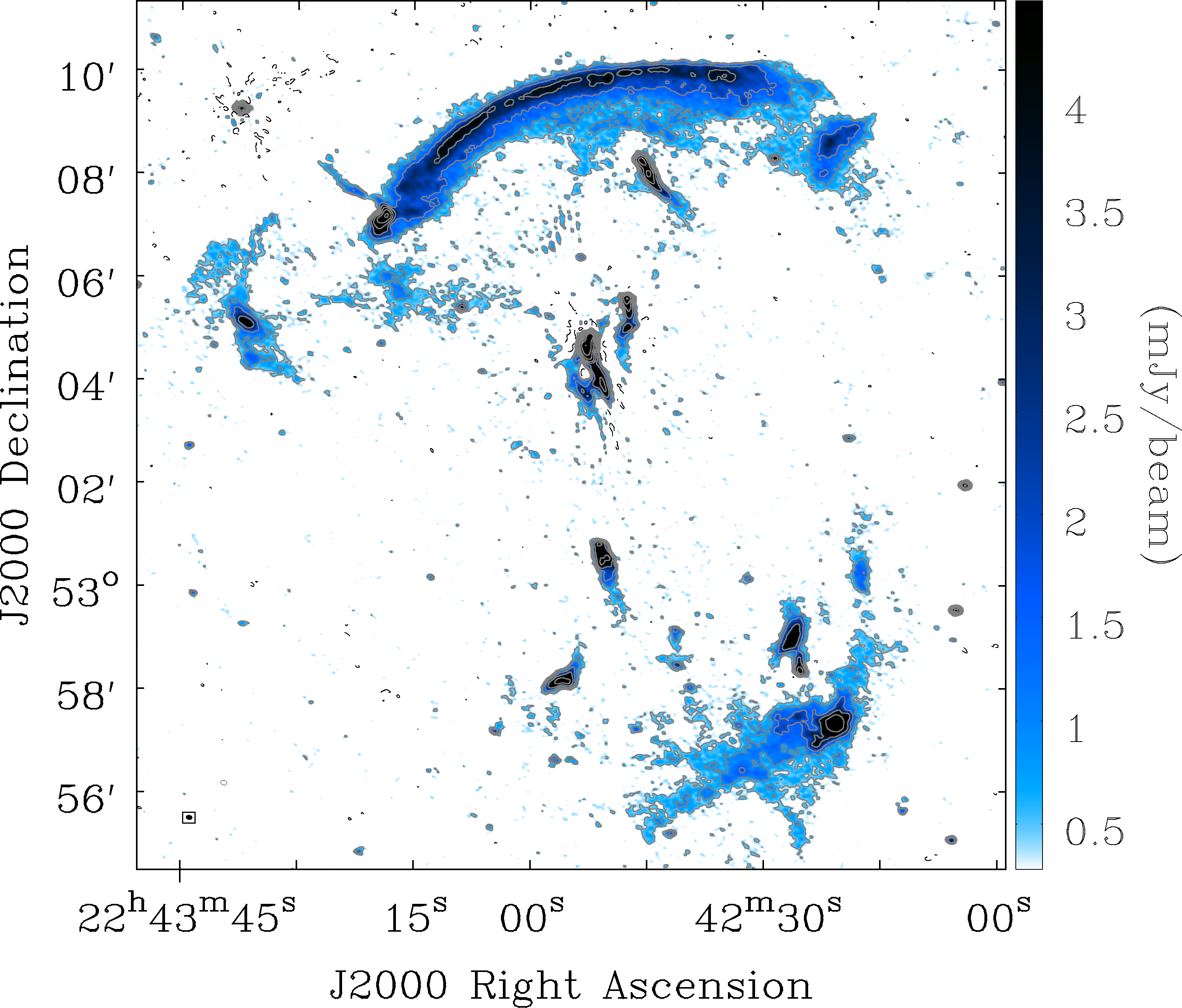}\caption{LOFAR total intensity high-resolution ($7.3\arcsec\times5.3\arcsec$,
bottom\textit{ }left corner) map of CIZA2242 and its contours levelled
at $[-3,3,\,6,\,12,\,24,\,48,\,96,\,192,\,384]\times\sigma_{\text{noise}}$,
$\sigma_{\text{noise}}=140\,\textrm{\ensuremath{\mu}Jy/beam}$. The
negative contours are black dashed lines. \label{fig:SS_Hres}}
\end{figure*}

\begin{figure}
\centering{}\includegraphics[clip,width=1\columnwidth]{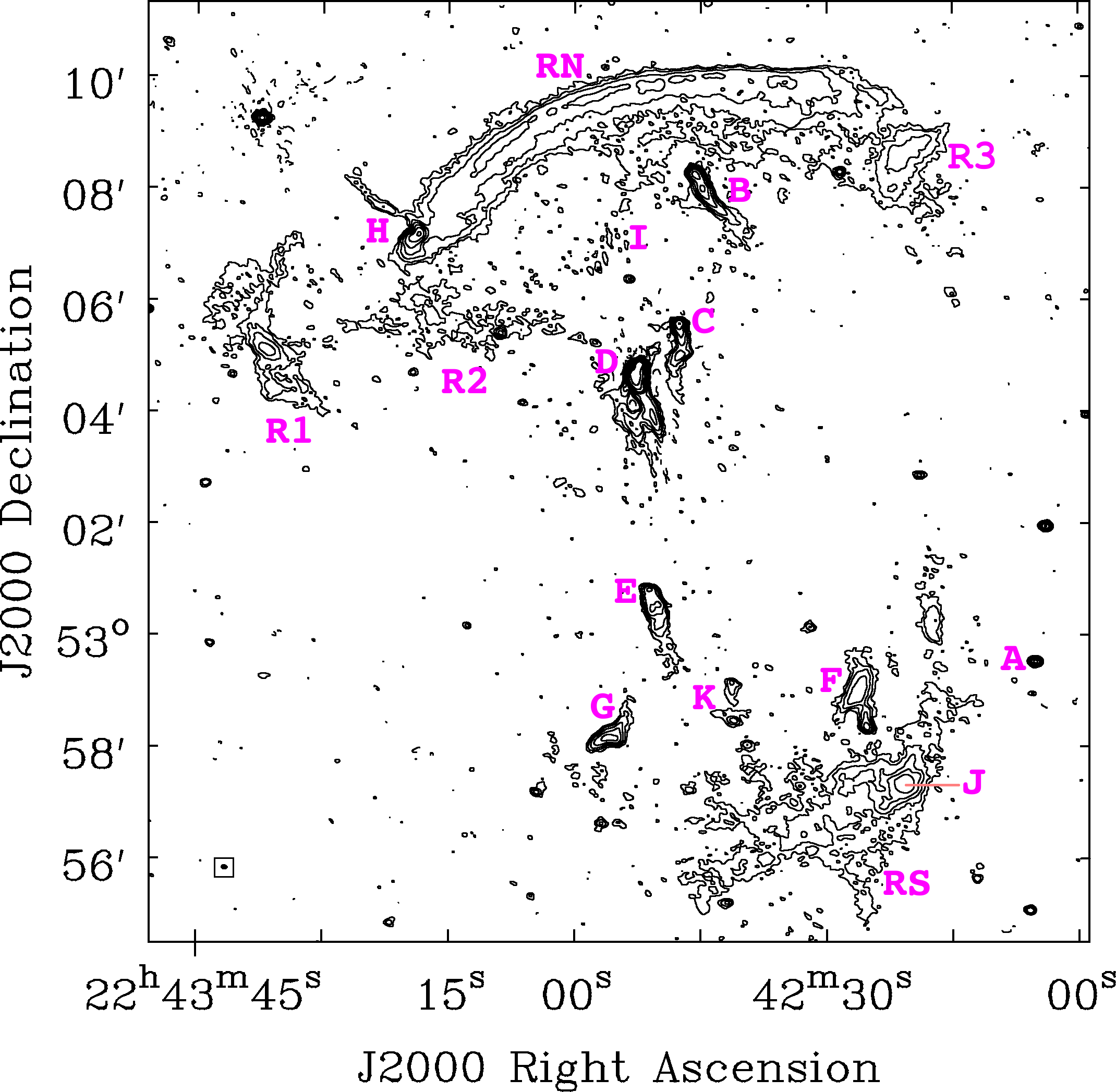}\caption{Source labels are adapted from \citealt{Stroe2013a}. We label the
patchy emission west of RN as R3. The contours are identical to those
in Fig. \ref{fig:SS_Hres}. \label{fig:SS_Hres_Labels}}
\end{figure}

\begin{figure*}
\centering{}\includegraphics[clip,width=0.7\textwidth]{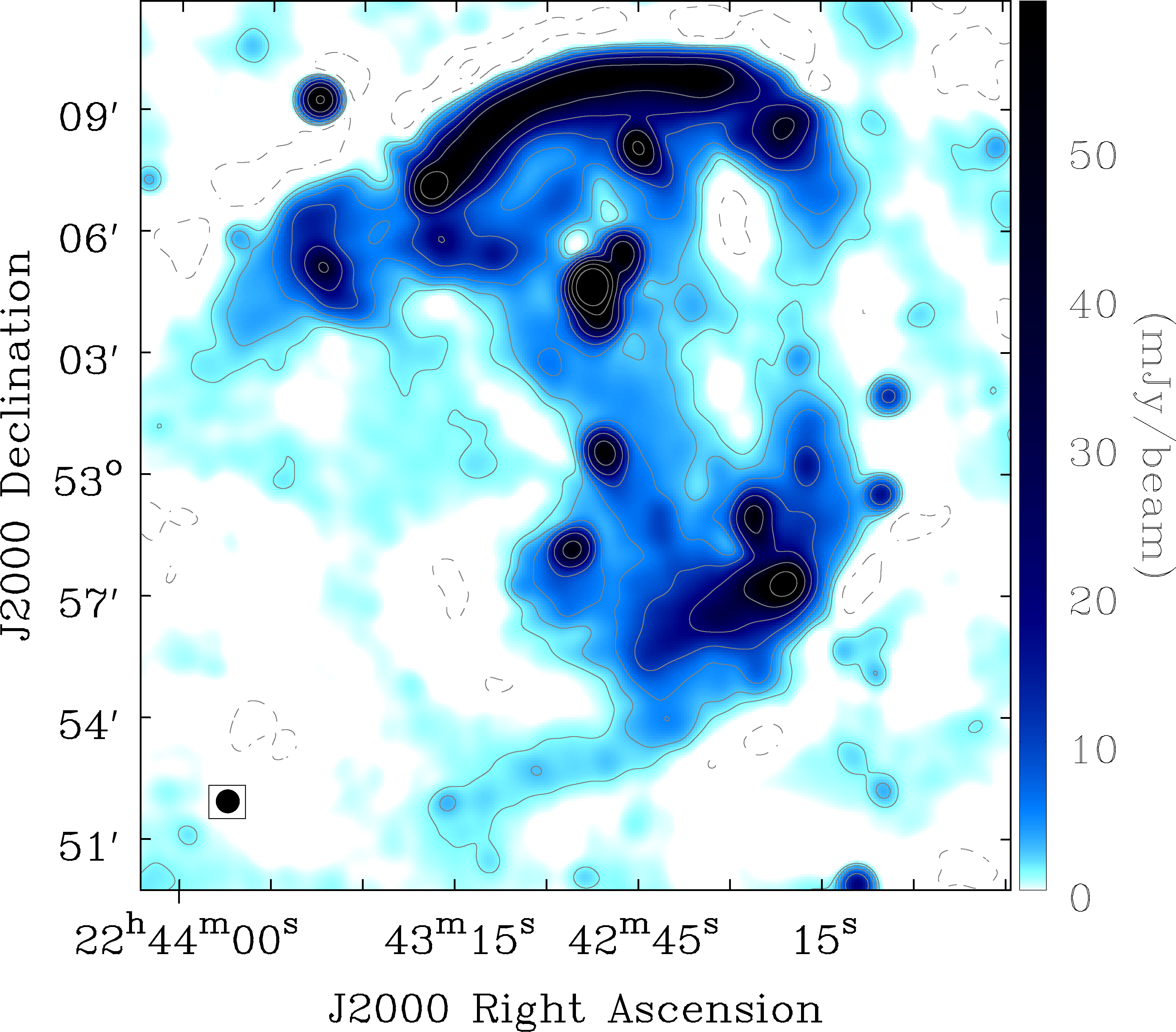}\caption{LOFAR total intensity low-resolution ($35\arcsec$, bottom\textit{
}left corner) image of CIZA2242. The radio contour levels are at $[-6,\,-3,\,3,\,6,\,12,\,24,\,48,\,96,\,192,\,384]\times\sigma_{\text{noise}}$
($\sigma_{\text{noise}}=430\,\textrm{\ensuremath{\mu}Jy/beam}$).
The negative contours are dashed lines. \label{fig:SS_Lres} }
\end{figure*}

\begin{figure}
\centering{}\includegraphics[clip,width=1\columnwidth]{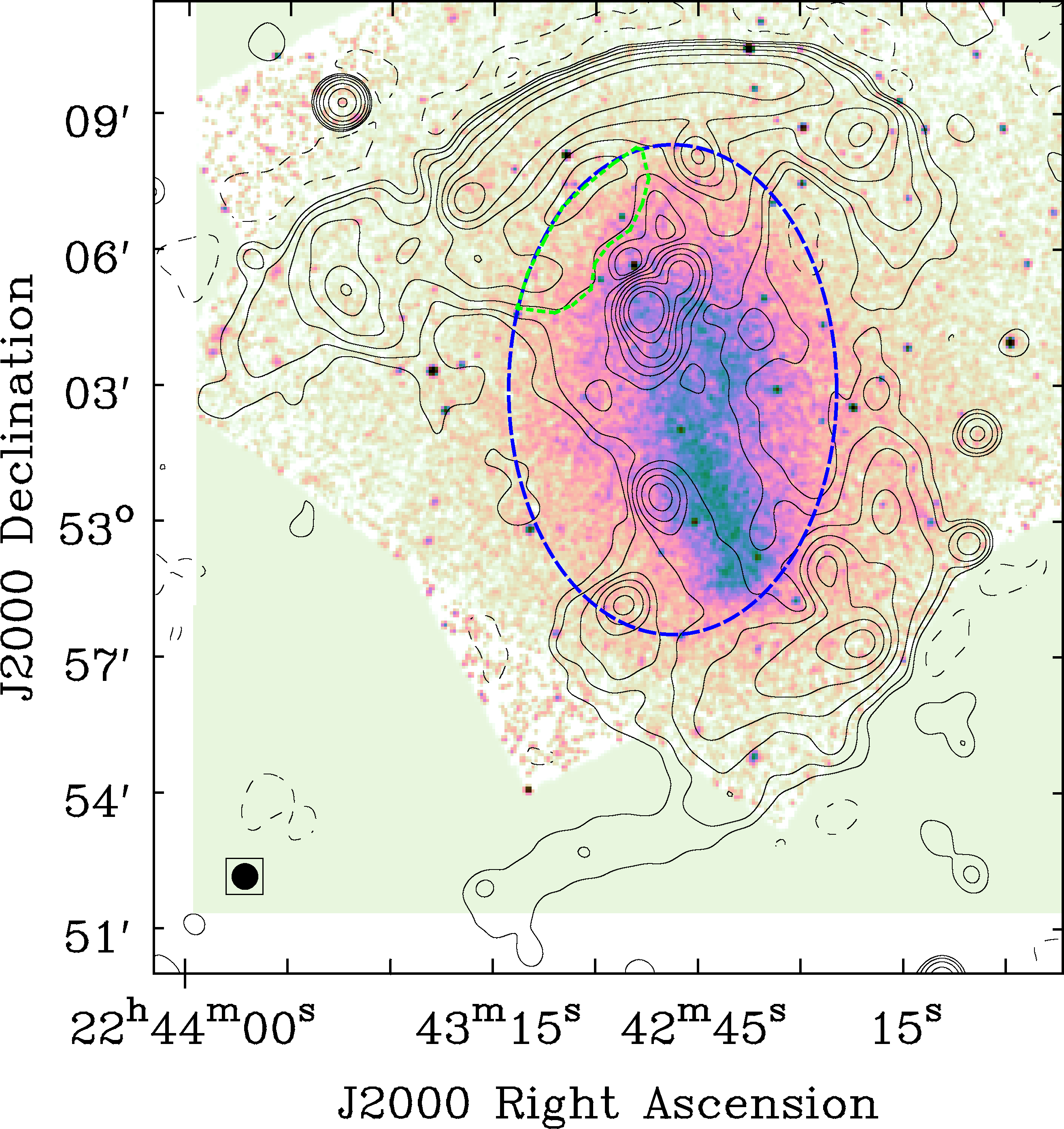}\caption{LOFAR  low-resolution ($35\arcsec$, bottom\textit{ }left corner)
contours of CIZA2242 on top of Chandra X-ray image in the $0.5-7.0\,\text{keV}$
band (being smoothed to $6\arcsec$ resolution, \citealt{Ogrean2014}).
The radio contour levels are identical to those in Fig. \ref{fig:SS_Lres}
The blue dashed elliptical region is approximately the halo emission
region. The green dotted region is over the source I. \label{fig:SS_Lres_Xray}}
\end{figure}

\begin{figure*}
\centering{}%
\noindent\begin{minipage}[t]{1\textwidth}%
\begin{center}
\includegraphics[clip,width=0.85\textwidth]{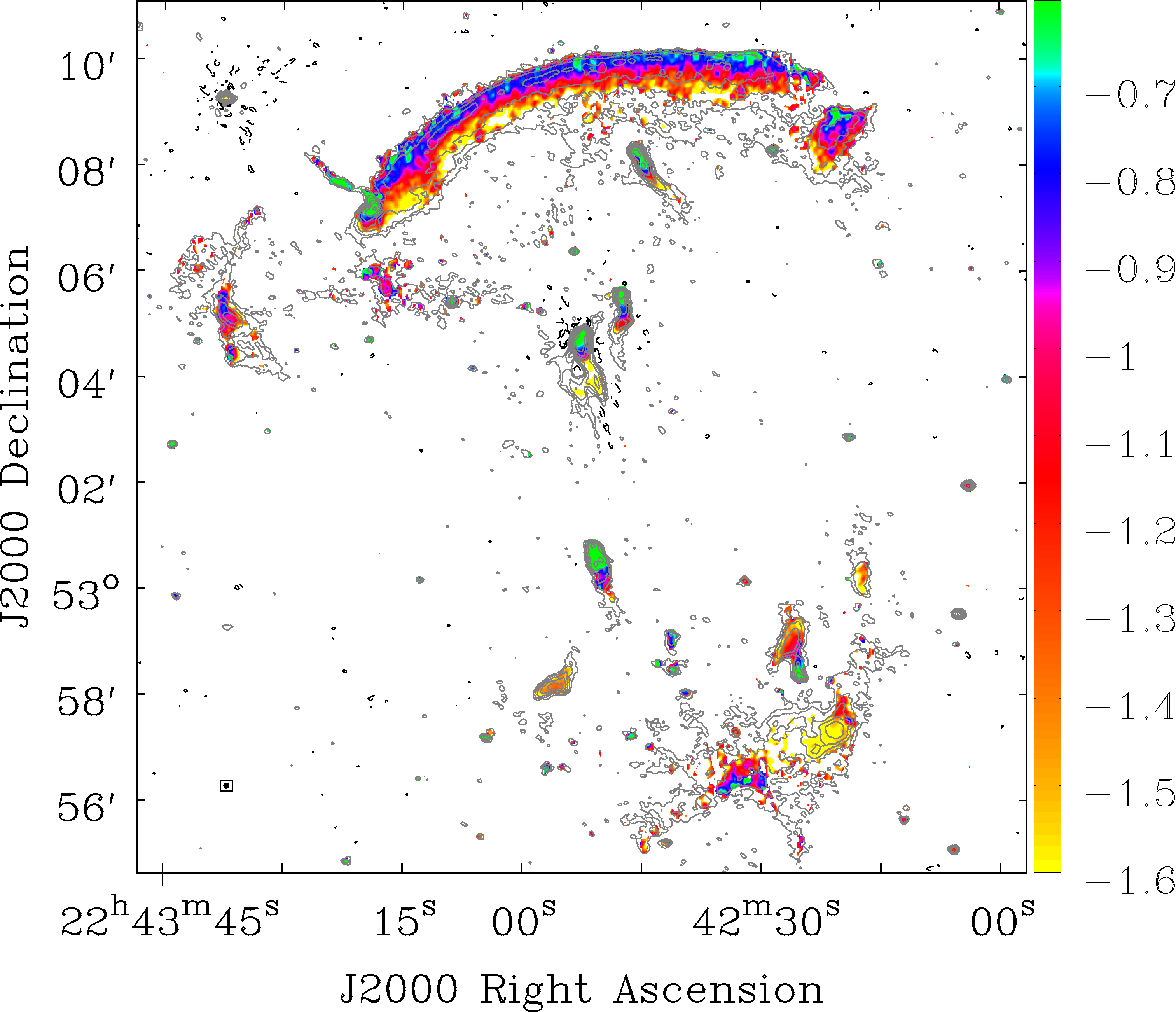}
\par\end{center}%
\end{minipage}\caption{A $6.5\arcsec-$resolution spectral index map of CIZA2242 from 145
to 608 MHz. The LOFAR contours are identical to those in Fig. \ref{fig:SS_Hres}.
The corresponding error map is shown in Fig. \ref{fig:SPX_Error_Map_Hres}.
\label{fig:SPX_HLres} }
\end{figure*}

In Fig. \ref{fig:SS_Hres}, we present our deep, high-resolution ($7.3\arcsec\times5.3\arcsec$)
LOFAR 145 MHz image of CIZA2242. The RMS noise reaches $140\,\textrm{\ensuremath{\mu}Jy/beam}$,
making this image one of the deepest, high-resolution, low-frequency
($<200\,\mbox{MHz}$) radio images of a galaxy cluster. The labelling
convention of \citet{Stroe2013a} is adopted and is presented in Fig.
\ref{fig:SS_Hres_Labels}. In Fig. \ref{fig:SS_Lres}, we show a
low-resolution ($35\arcsec$) LOFAR image. The low-resolution contours
are plotted over a Chandra X-ray image (smoothed to $6\arcsec$ resolution
using a Gaussian kernel, \citealt{Ogrean2014}) in Fig. \ref{fig:SS_Lres_Xray}.
 In Fig. \ref{fig:SPX_HLres}, we show our high-resolution ($6.5\arcsec$)
spectral index map from 145  to 608 MHz.

\subsection{Northern relic \label{subsec:RN}}

\begin{figure}
\centering{}\includegraphics[clip,width=1\columnwidth]{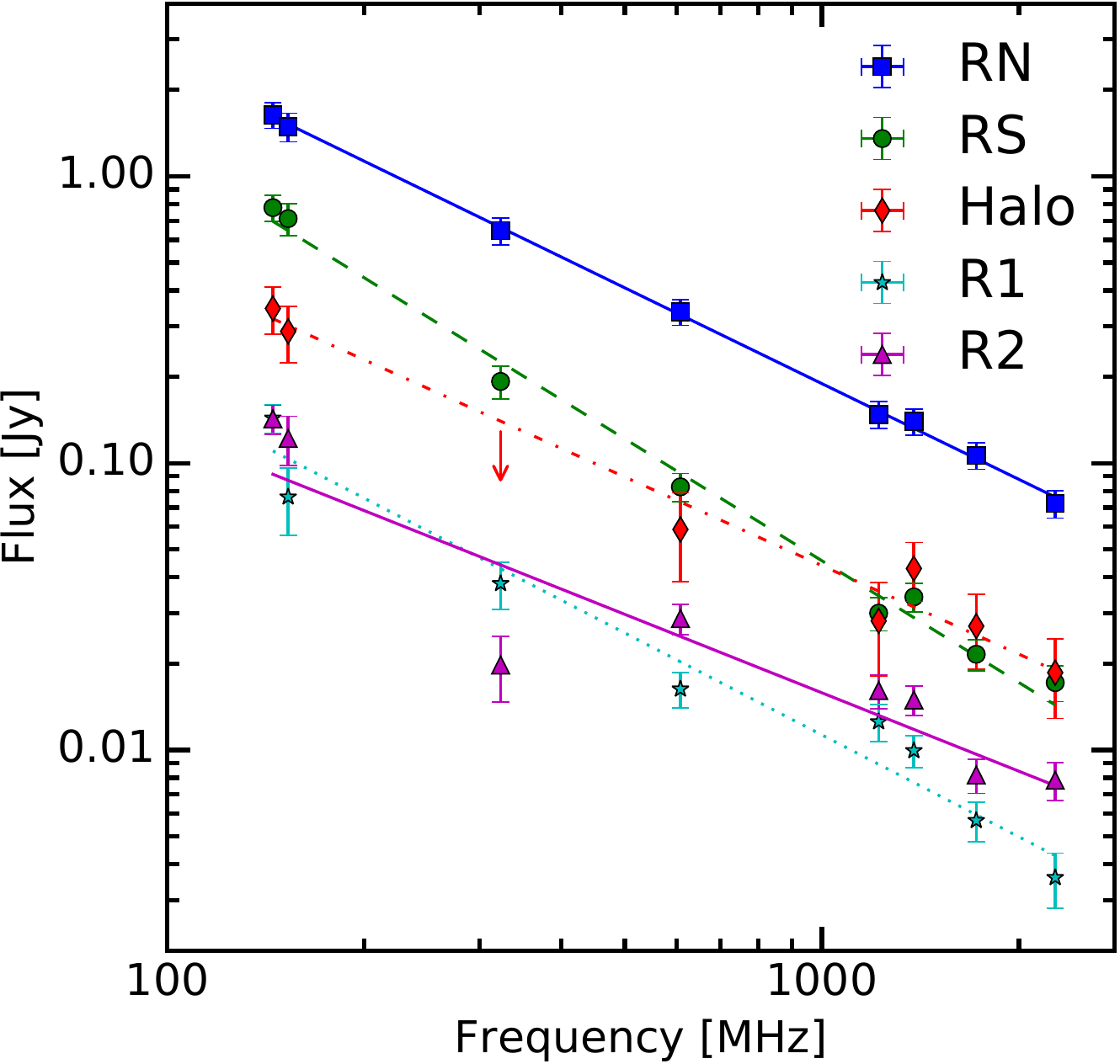}\caption{Integrated radio spectra of RN, RS, the halo, R1 and R2 from 145 MHz
to 2.3 GHz including the LOFAR data points at the lowest frequency
145 MHz. The red downwards pointing arrow is the upper limit derived
from the GMRT 323 MHz observations. The substantial scatter around the
lines of best fit for the halo, R1 and R2 (e.g. compared with RN and
RS) could be caused by the low signal to noise of the detections and
the difficulty in precisely imaging large faint diffuse sources. Given
the comparatively high surface brightness sensitivity of LOFAR, it
is possible that the spectral indices for these sources is artificially
steepened. The weighted best-fit integrated spectral indices are listed
in Table \ref{tab:Int_Flux_Spx}.  The integrated fluxes for the relics and halo are given in Table \ref{tab:Int_fluxes}. \label{fig:Spx_Int_RN_RS_Halo}
}
\end{figure}

The northern relic (RN) is nicely mapped out in our LOFAR image (Fig.
\ref{fig:SS_Hres}). Its morphology shows a familiar arc-like feature
with a projected linear size of $1.9\,\mbox{Mpc}\times230\,\mbox{kpc}$
(measured at $3\sigma_{\text{noise}}$ contours). The length of RN
increases to $2.1\,\text{Mpc}$ when measured in the $35\arcsec$
map (Fig. \ref{fig:SS_Lres}). Its surface brightness has a sharp
edge on the northern side and a more gradual decline on the southern
side with additional diffuse emission north of source B. The integrated
flux of RN (including the patchy emission in the west, source R3 in
Fig. \ref{fig:SS_Hres_Labels}) is measured to be $1548.2\pm4.6\,\mbox{mJy}$
within the $\geqslant3\,\sigma_{\text{noise}}$ region. The flux increases
by $4.6\%$ and $8.3\%$ for the $\geqslant2\sigma_{\text{noise}}$
and $\geqslant1\sigma_{\text{noise}}$ regions, respectively. The
spectral index map between 145 and 608 MHz (Fig. \ref{fig:SPX_HLres})
shows a clear steepening from the north towards the cluster centre,
ranging roughly from $-0.80$ to $-1.40$. In Fig. \ref{fig:Spx_Int_RN_RS_Halo},
we plot the integrated fluxes of RN between 145 MHz and 2.3 GHz which
were calculated within the LOFAR $\geqslant3\sigma_{\text{noise}}$
region. The spectral index obtained from  a weighted least-squares
fitting of a power-law function to the RN fluxes at eight frequencies
is $-1.11\pm0.04$, which is consistent with the previous values of
$-1.08\pm0.05$ \citep{VanWeeren2010a} and $-1.06\pm0.04$ \citep{Stroe2013a}.

Towards the western side of RN, the main relic connects with source
R3 via a faint bridge (a $3\sigma_{\text{noise}}$ detection). Towards
the eastern side, RN is attached to source H, which is much brighter
than other regions of the relic (the peak brightness is $30\,\mbox{mJy/beam}$,
compared with a typical brightness $4.5\,\mbox{mJy/beam}$ in RN).
 The northern emission associated with source H has the expected morphology
for an AGN.

In the post-shock central region of RN, an excess of emission was
detected at a significance of up to $6\sigma_{\text{noise}}$ in front
of source B. This emission  has an arc-like shape with a projected
linear size of $\sim135\,\text{kpc}\times500\,\mbox{kpc}$ within
the $3\sigma_{\text{noise}}$ contours (Fig. \ref{fig:SS_Hres}).
Interestingly, this feature was not detected in the post-shock eastern
region of RN, where no tailed AGN  are observed. The excess emission
 is more visible in the surface brightness profiles along the width
of RN for the central and eastern regions in Fig. \ref{fig:Post-Shock_Profiles}.
Since the excess emission is located in front of source B and has
a shock-like morphology, we speculate that this excess emission  could
be a bow shock generated by an interaction between the tailed AGN
(source B) and the downstream relativistic electrons. We will present
further analysis of this speculation in an  upcoming paper.
\begin{center}
\begin{figure}
\centering{}
\includegraphics[width=1\columnwidth]{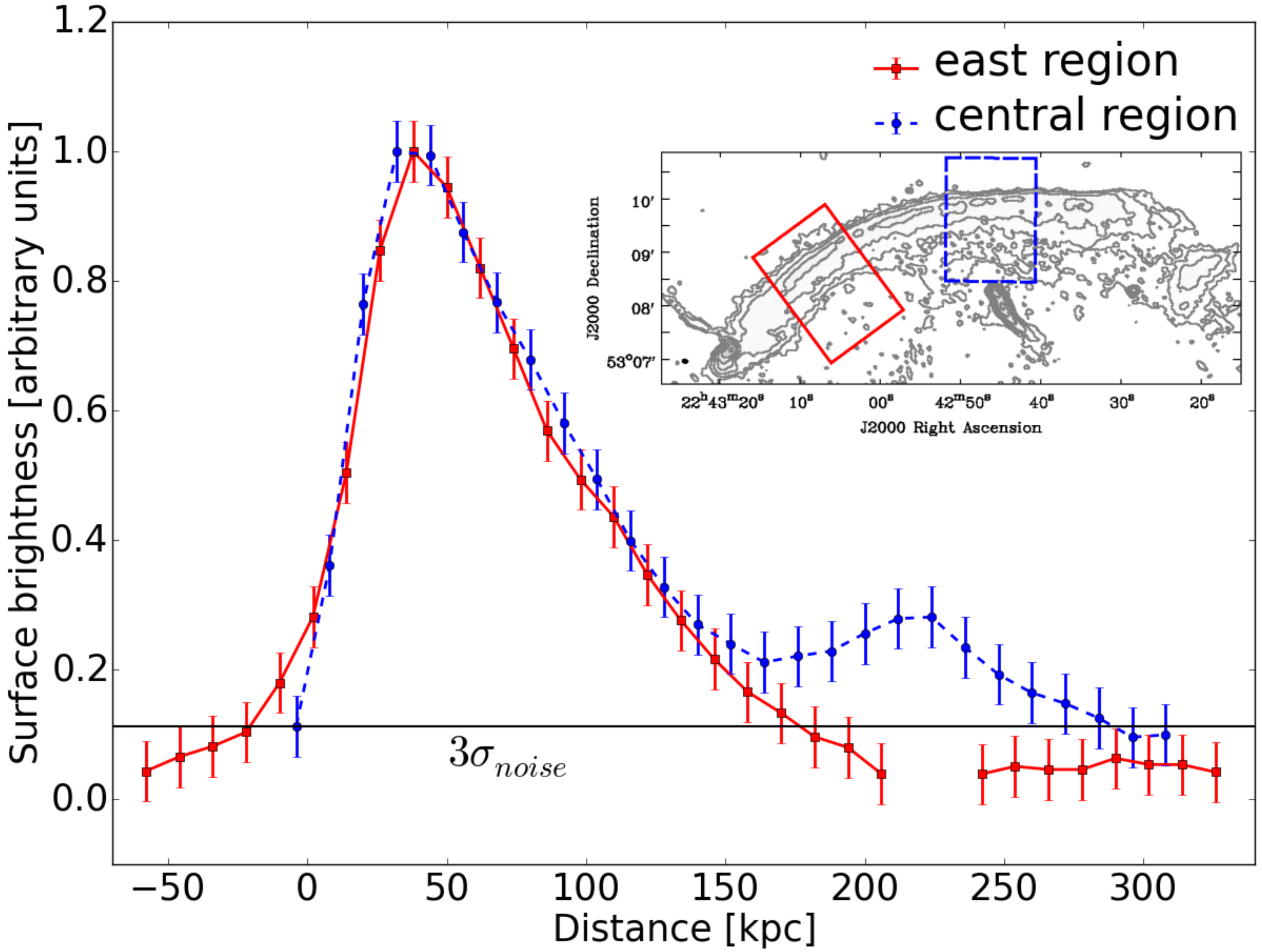}\caption{Surface brightness profiles (normalised) along the width of RN (in
north-south direction) for the central (blue dashed) and east (red
solid) regions (overlaid image). The data ($\geqslant1\sigma_{\text{noise}}$)
is averaged within $4\arcsec$ along the profile. \label{fig:Post-Shock_Profiles}
}
\end{figure}
\par\end{center}

\subsection{Southern relic \label{subsec:RS}}

The southern relic (RS) is detected in our LOFAR map in Fig. \ref{fig:SS_Hres}
with a peak signal to noise of $\sim20$. The distance between the
outer edges of RN and RS is measured to be $2.8\,\text{Mpc}$ or $3.2\,\text{Mpc}$
in the $6.5\arcsec$ or $35\arcsec$ maps, respectively. The width
of RS is wider in the central region with a size of $\sim270\,\text{kpc}$
within the $\geqslant3\sigma_{\text{noise}}$ region (Fig. \ref{fig:SS_Hres}).
Within this central area the region labelled J covers $\sim200\,\text{kpc}$
in radius and has substantially higher (threefold) surface brightness
than the rest of the relic but no obvious counterpart in the optical
data (see Fig. A.1. in \citealt{Stroe2013a}). 

The  emission in the region of RS includes  newly detected faint diffuse
emission in the south-east region (Fig. \ref{fig:SS_Lres_Xray} or
Fig. \ref{fig:SS_lowres_optical_radio}). The size of RS in projection
covers a maximum linear distance of $\sim1.4\,\text{Mpc}$ as measured
within the $3\sigma_{\text{noise}}$ contours in the $6.5\arcsec$
map. But its length when measured in the $35\arcsec$ map in Fig.
\ref{fig:SS_Lres_Xray} significantly increases to $\sim2\,\text{Mpc}$
 or $\sim3\,\text{Mpc}$  when excluding or including the south-east
emission. Within this new south-east  region of emission, there are
four optical sources which correspond to peaks in the radio emission
as shown in Fig. \ref{fig:SS_lowres_optical_radio}. This  faint emission
could be the result of  a collection of compact sources or it could
be an in-falling filament at the cluster redshift. Unfortunately,
we were unable to constrain the redshifts for the optical sources
with the existing optical data. Additionally, faint diffuse emission
is observed to extend south-westwards from the central region of RS
 (Fig. \ref{fig:SS_Hres}). 

\begin{figure*}
\centering{}%
\noindent\begin{minipage}[t]{1\columnwidth}%
\begin{center}
\includegraphics[clip,width=0.74\columnwidth]{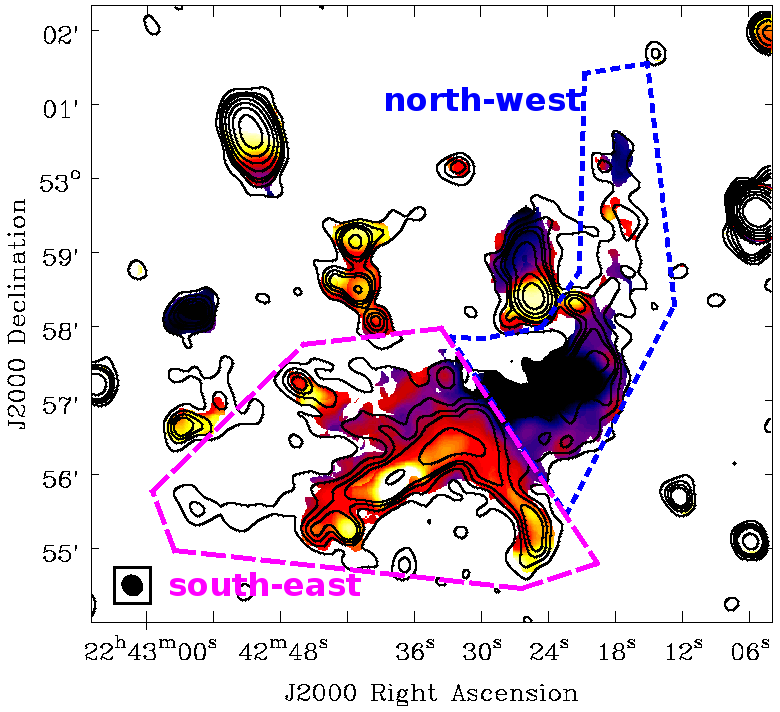}
\par\end{center}%
\end{minipage}%
\noindent\begin{minipage}[t]{1\columnwidth}%
\begin{center}
\includegraphics[width=1\columnwidth]{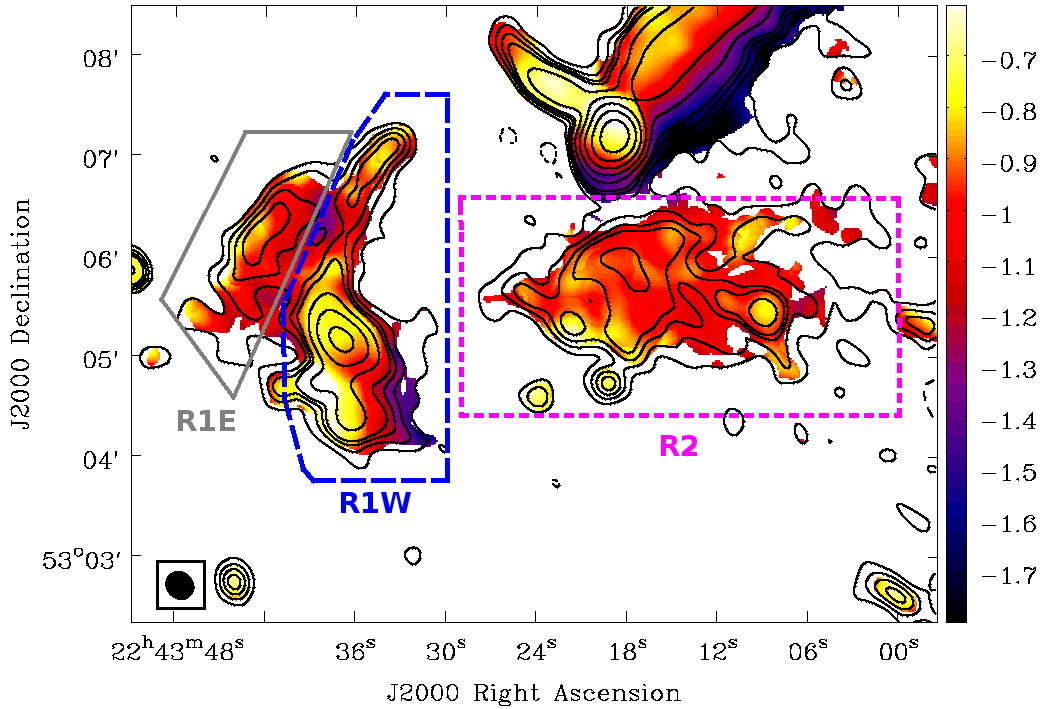}
\par\end{center}%
\end{minipage}\caption{Zoom-in mid-resolution ($18\arcsec\times16\arcsec$) spectral index
maps of RS (left) and the eastern relics (right) from 145 MHz to 2.3
GHz (eight frequencies). The overlaid WSRT 1.4 GHz contours at the
same resolution are levelled at $[3,6,9,12,24,48,96,192,384,768]\times\sigma_{\text{noise}}$($\sigma_{\text{noise}}=31.2\,\text{\ensuremath{\mu}Jy/beam}$).
\label{fig:SPX_RS_R1_R2} }
\end{figure*}

The zoom-in spectral index map of RS (eight frequencies from 145 MHz
to 2.3 GHz) that is shown in Fig. \ref{fig:SPX_RS_R1_R2} shows steepening
towards the cluster centre. The spectral index drops approximately
from $-0.85$ to $-1.40$ in the south-east region and from $-1.35$
to $-1.70$ in the north-west region. Similar trends are seen in the
$145-608\,\text{MHz}$ spectral index map in Fig. \ref{fig:SPX_HLres}.
The north-west region of RS (Fig. \ref{fig:SPX_RS_R1_R2}) is dominated
by diffuse emission in region J which has a very steep mean spectral
index of $-2.0$. The region of flat spectrum in the south-east part
(Fig. \ref{fig:SPX_RS_R1_R2}, left) has an L shape appearance and
a steep spectrum of mean value of $-0.93\pm0.10$. The west region
of RS (Fig. \ref{fig:SPX_RS_R1_R2}, left) has an arc-like shape with
an average spectral index of $-1.30\pm0.04$ (excluding region J).
We estimated the integrated spectral index for RS (within the LOFAR
$\geqslant3\sigma_{\text{noise}}$ region, including region J) to
be $-1.41\pm0.05$ (Fig. \ref{fig:Spx_Int_RN_RS_Halo}), which is
steeper than the value of $-1.29\pm0.04$ that was reported in \citet{Stroe2013a}.

\subsection{Eastern relics \label{subsec:RE}}

East of the cluster centre, slightly south, from RN (Fig. \ref{fig:SS_Hres})
are the relics R1 and R2, which were previously identified in \citet{Stroe2013a}.
R1 consists of a thin narrow region. It is $\sim630\,\mbox{kpc}$
long and oriented in the north-south direction. R1 has a bright region
in the middle and patchy emission in the north-east. In Fig. \ref{fig:SPX_RS_R1_R2}
we show the eight-frequency spectral index distribution ($18\arcsec\times16\arcsec$)
of the eastern relics. Based on morphology and spectral index, we
divided R1 into two regions: R1E in the north-east and R1W in the
north-west (see Fig. \ref{fig:SPX_RS_R1_R2}). R1W has an arc-like
morphology and shows spectral index steepening from approximately
$-0.70$ to $-1.40$ towards the centre. The spectral index of R1E
drops from approximately $-0.75$ to $-1.20$ in the same direction.

R2 has a physical size of $670\,\mbox{kpc}\times270\,\mbox{kpc}$
with the major axis in the west-east direction (Fig. \ref{fig:SPX_RS_R1_R2},
right). Its spectral index between 145 MHz and 2.3 GHz remains approximately
constant across its structure with a mean of $-0.95\pm0.08$.

The integrated fluxes over the $\geqslant3\sigma_{\text{noise}}$
region in the LOFAR $18\arcsec\times16\arcsec$ image after masking
out the compact sources were estimated to be $143.5\pm8.3\text{\,mJy}$
and $142.5\pm8.1\,\text{mJy}$ for R1 and R2, respectively. The integrated
spectral indices between $145\,\text{MHz}$ and $2.3\,\text{GHz}$
are $-1.18\pm0.06$ for R1 and $-0.91\pm0.06$ for R2.

\subsection{Radio halo \label{subsec:RH}}

The halo was discovered by \citet{VanWeeren2010a} with WSRT observations
at 1.4 GHz. Part of the emission was also seen in the GMRT 153 MHz
image that was presented in \citet{Stroe2013a}. However, its flux
and spectral properties have not been well studied. With our deep LOFAR
image, we are able to characterise the radio halo as the large number
of short baselines give LOFAR excellent sensitivity to cluster-scale
emission. In Fig. \ref{fig:SS_Lres}, we show a low-resolution ($35\arcsec$)
image of CIZA2242.  The image shows diffuse emission connecting
RN and RS with a significance up to $9\sigma_{\text{noise}}$ ($\sigma_{\text{noise}}=430\,\textrm{\ensuremath{\mu}Jy/beam}$).
The emission covers an area of $1.8\,\text{Mpc}\times830\,\text{kpc}$
with its major axis elongated in the north-south direction broadly
following the Chandra X-ray emission that was mapped by \citet{Ogrean2014}.

The $145\,\text{MHz}$ flux of the radio halo was estimated from the
LOFAR data in an elliptical region that was selected to cover the
Chandra X-ray emission (Fig. \ref{fig:SS_Lres_Xray}). However, this
region also hosts radio galaxies (i.e. source B, C, D, E, F, G, K)
and diffuse emission from source I. We attempted to remove these contaminating
sources from the halo flux estimation in two steps. In the first step,
models of the radio galaxies were subtracted from the uv data. To
obtain the radio galaxy models we created a LOFAR image using parameters
that are similar to those used for the high-resolution ($6.5\arcsec$)
image, except the inner uv cut was set to $0.4\,\text{k\ensuremath{\lambda}}$
instead of $0.2\,\text{k\ensuremath{\lambda}}$ (see Table \ref{tab:Imaging_Parameters}
for the imaging parameters). Here the inner uv cut was used to filter
out the large-scale emission from the halo to leave only the radio
galaxies. The CLEAN components of these radio galaxies were subtracted
from the uv data which were then imaged and smoothed to obtain a $35\arcsec$
image. The radio halo flux $S_{1}$ in the $\geqslant3\sigma_{\text{noise}}$
elliptical region (blue dashed in Fig. \ref{fig:SS_Lres_Xray}), with
source I masked, was measured from this $35\arcsec$ image. In the
second step, the radio halo flux $S_{2}$ in the source I region (green
dotted in Fig. \ref{fig:SS_Lres_Xray}) was estimated by extrapolating
the halo flux $S_{1}$ using a scaling factor proportional to ratio
of the areas (i.e. $\nicefrac{\text{Area}_{S_{2}}}{\text{Area}_{S_{1}}}=0.085$).
We did not subtract source I from the uv data due to it having large-scale
diffuse emission which is difficult to disentangle from the halo emission.
Finally, the total halo flux ($S_{h}=S_{1}+S_{2}$) was estimated
to be $S_{h}^{\text{145\,MHz}}=346\pm64\,\text{mJy}$. Where we calculated
the total error following \citet{Cassano2013a} and took into account
the flux scale uncertainty (i.e. $10\%$), image noise over the halo
area and source subtraction uncertainty (i.e. $5\%$ of the total
flux of the subtracted radio galaxies); additionally, we added an
estimate of the uncertainty associated with the extrapolation of the
halo flux in the source I region (i.e. assuming $10\%$). The source
subtraction error of $5\%$ was estimated from the ratio of the post
source subtraction residuals to the pre source subtraction flux of
a nearby compact source (i.e. at RA=22:432:37, Dec=+53.09.16). Another
uncertainty in the integrated flux is the area used for the integration.
To assess the dependence on this we estimated the integrated radio
flux within regions bounded by $\geqslant2\sigma_{\text{noise}}$
and $\geqslant1\sigma_{\text{noise}}$ within the elliptical region.
We found the total halo flux at $145\,\text{MHz}$ remained approximately
stable ($S_{h}^{145\,\text{MHz}}=362\pm65\,\text{mJy}$ for $\geqslant2\sigma_{\text{noise}}$
or $S_{h}^{145\,\text{MHz}}=371\pm66\,\text{mJy}$ for $\geqslant1\sigma_{\text{noise}}$).

Following a similar procedure we estimated the halo flux in the GMRT/WSRT
data sets within the same region used for the LOFAR data. Unfortunately, due to the depth of the GMRT/WSRT observations,
the halo emission was only partly detected in the GMRT 153, 608 MHz
and WSRT 1.2, 1.4, 1.7, 2.3 GHz maps and undetected in the GMRT 323
MHz map. This may induce systematics in the spectral index estimate that deserve sensitive observations at higher frequencies in the future. In this paper we attempt our best following the approaches adopted in previous papers \citep[e.g.][]{VanWeeren2016b, Stroe2013a}. First we used a common uv range (i.e. $0.2-50\,\text{k}\lambda$) and Briggs weighting (robust=0.5) when making the $35\arcsec$ images  (see Subsec. \ref{subsec:Spectral_Index_maps}). We also added an absolute flux scale uncertainty of $10\%$ to mitigate the impact on our conclusions due to possible missing flux in some observations.  Due to the large uncertainties that are propagated in the
above procedure (i.e. flux scale, image noise, source subtraction
and extrapolation), we only created source subtracted low-resolution
($35\arcsec$) images. From these we estimated the inverse-variance weighted integrated
spectrum which is plotted in Fig. \ref{fig:Spx_Int_RN_RS_Halo}. Using
the combination of LOFAR, GMRT and WSRT data sets, the integrated
spectral index from $145\,\text{MHz}$ to $2.3\,\text{GHz}$ of the
radio halo was estimated to be $-1.03\pm0.09$ and was found to remain
approximately constant for different $\sigma_{\text{noise}}$ cuts
on the area of integration (i.e. $1-3\sigma_{\text{noise}}$). 

\begin{table}
\caption{Integrated spectral indices between
145 MHz and 2.3 GHz. \label{tab:Int_Flux_Spx}}

\begin{centering}
\begin{tabular}{ccc}
 \hline
Source &  $\alpha_{\text{int}}^{a}$ & $\alpha_{\text{int}}^{b}$\tabularnewline
 \hline
RN &  $-1.11\pm0.04$ & $-1.06\pm0.04$\tabularnewline
RS &  $-1.41\pm0.05$ & $-1.29\pm0.04$\tabularnewline
R1 &  $-1.18\pm0.06^{c}$ & $-0.74\pm0.07$\tabularnewline
R2 &  $-0.91\pm0.06$ & $-0.90\pm0.06$\tabularnewline
Halo  & $-1.03\pm0.09^{d}$ & N/A\tabularnewline
 \hline 
\end{tabular}\tabularnewline
$^{a}$: a flux scale uncertainty of $10\%$; $^{b}$: \citealt{Stroe2013a}; $^{c}$: compact source was masked; $^{d}$: compact source subtraction error of  5\% and an uncertainty of 10\% from the extrapolation of  the halo emission in the source I region were added. 
\par\end{centering}
\end{table}

\subsection{Tailed radio galaxies}

Our LOFAR maps show that there are at least six tailed radio galaxies
in CIZA2242 (i.e. source B, C, D, E, F, G in Fig. \ref{fig:SS_Hres}),
a remarkably large number for any cluster. We mentioned in Sec. \ref{sec:Introduction}
 that direct acceleration of thermal electrons by merger shocks is
unable to explain the observed spectra and the efficiencies of electron
acceleration for a number of relics (e.g. \citealt{Akamatsu2015,Vazza2015,VanWeeren2016b,Botteon2016}).
These problems could be solved with shock re-acceleration of fossil
relativistic electrons. Tailed radio galaxies (e.g. \citealt{Miley1972})
are obvious reservoirs of such electrons, after activity in their
nuclei has ceased or become weak. Indeed, some radio tails, such as
3C 129 (e.g. \citealt{Lane2002}) have Mpc-scale lengths and morphologies
that bear striking resemblance to those of relics. Recent studies
show evidence for particle re-acceleration of fossil electrons from
radio galaxies in the radio relics of PLCKG287.0+32.9 (\citealt{Bonafede2014})
and Abell 3411-3412 (\citealt{VanWeeren2017}). In the case of CIZA2242
there are bright radio galaxies located at the eastern extremity of
RN (i.e. source H) and at the brightest central region of RS (i.e.
source J). Whilst it remains unclear whether these radio galaxies
are related to RN a scenario where fossil plasma could contribute
to the emission in RN was previously discussed in \citet{shimwell2015}. 

The combined systematic study of relics and tailed radio galaxies
with radio sky surveys (e.g. \citealt{Norris2011,Rottgering2011,Shimwell2017})
may shed light on the role of tailed radio galaxies in the formation
of radio relics, in particular in the case that radio relics originate
from re-acceleration of fossil relativistic electrons and rather than
from the acceleration of thermal plasma.

\section{Discussion}
\label{sec:Discussion}

CIZA2242 is one of the best studied merging clusters. In the radio
band \citet{Stroe2013a} carried out a detailed study of the cluster
sources with  GMRT/WSRT at  frequencies ranging from 153 MHz to 2.3
GHz. Our LOFAR observations have produced higher-sensitivity, deeper
images than the existing images at equivalent frequencies. In this
section, we only discuss the new results  obtained using the LOFAR
images. We discuss below the discrepancy in the Mach numbers of RN
and RS that were derived from X-ray and radio data, the characteristics
of the post shock emission in RN, the eastern relic R1, and the radio
halo.

\subsection{Radio spectrum derived Mach numbers \label{subsec:Mach_Number}}

The underlying particle-acceleration physics of shock waves closely
relates to the shock Mach number and the observed spectra (e.g. \citealt{Blandford1987,Donnert2016,Kang2016b}):

\begin{equation}
\mathcal{M}=\sqrt{\frac{2\alpha_{\text{inj}}-3}{2\alpha_{\text{inj}}+1}},\label{eq:Mach}
\end{equation}
where the injection spectral index $\alpha_{\text{inj}}$ is related
to the power-law energy distribution of relativistic electrons ($dN(p)/dp\propto p^{-\delta_{\text{inj}}}$,
where $dN(p)$ is the electron number within momentum range $p$ and
$p+dp$) via the relation $\alpha_{\text{inj}}=-(\delta_{\text{inj}}-1)/2$.
For a simple planar shock model \citep{Ginzburg1969} the injection
index $\alpha_{\text{inj}}$ is flatter than the volume-integrated
spectral index $\alpha_{\text{int}}$,
\begin{equation}
\alpha_{\text{inj}}=\alpha_{\text{int}}+0.5.\label{eq:Inj_Int_Approx}
\end{equation}
However,  recent DSA test-particle simulations by \citet{Kang2015a,Kang2015b}
indicate that this approximation (Eq. \ref{eq:Inj_Int_Approx}) breaks
down for spherically expanding shock waves, due to shock compression
and the injection electron flux gradually decreasing as the shock
speed reduces in time. In the following Subsec. \ref{subsec:M_SPX_Int}
and \ref{subsec:M_SPX_Map} we will estimate Mach numbers for the
northern and southern shocks using the injection spectral indices
that are measured from integrated spectra and resolved spectral index
maps.

\subsubsection{Mach numbers from integrated spectra \label{subsec:M_SPX_Int}}

Using  LOFAR data, together with existing radio data (\citealt{VanWeeren2010a,Stroe2013a}),
we measured volume-integrated spectral indices of $\alpha_{\text{int}}^{n}=-1.11\pm0.04$
for RN and of $\alpha_{\text{int}}^{s}=-1.41\pm0.05$ for RS. From
that we derived injection indices of $\alpha_{\text{inj}}^{n}=-0.61\pm0.04$
and $\alpha_{\text{inj}}^{s}=-0.91\pm0.05$ (Eq. \ref{eq:Inj_Int_Approx})
and Mach numbers of $\mathcal{M}_{n}=4.4_{-0.6}^{+1.1}$ and $\mathcal{M}_{s}=2.4\pm0.1$.
For $\mathcal{M}_{n}$ our result is consistent with the findings
of \citet{Stroe2013a} ($4.58\pm{1.09}$, using colour-colour
plots) and \citet{VanWeeren2010a} ($4.6_{-0.9}^{+1.3}$, using a
$0.68-2.3\,\text{GHz}$ spectral index map). However, $\mathcal{M}_{s}$
is significantly smaller than that estimated in \citet{Stroe2013a}
($2.80\pm0.19$). Furthermore, there remains a discrepancy between
the Mach numbers obtained from the integrated spectral index and those
derived from X-ray data or spectral age modelling of radio data (\citealt{Stroe2014g,Akamatsu2015})
(see Table \ref{tab:Spx_Mach}).

\subsubsection{Mach numbers from spectral index maps \label{subsec:M_SPX_Map}}

An alternative way to obtain the injection index is to measure it
directly from the shock front region in high resolution spectral index
maps. Unfortunately , the precise thickness of the region in which
the relativistic electrons are  (re-)accelerated by the shock is unknown.
However, given that the bulk velocity is approximately $905\,\text{km/s}$
in the downstream region (\citealt{Stroe2014g}), the travel time
for the electrons to cross a distance equivalent to the beam size
of $6.5\arcsec$ ($22.5\,\text{Myrs}$) is $4-10$ times shorter than
their estimated lifetime of $90-220\,\text{Myrs}$ (assuming the relativistic
electrons are in the magnetic field $\sim\text{\ensuremath{\mu}G}$
and observed at the frequencies $150-610\,\text{MHz}$, see e.g. \citealt{Donnert2016}).
Therefore, the shock (re-)accelerated relativistic electrons are not
likely to lose a significant amount of their energy along the distance
corresponding to the synthesized beam size; and the injection spectral
index can be approximated from measurements at the shock front.

\begin{figure*}
\centering{}%
\begin{minipage}[t]{0.42\textwidth}%
\begin{center}
\includegraphics[width=1\textwidth]{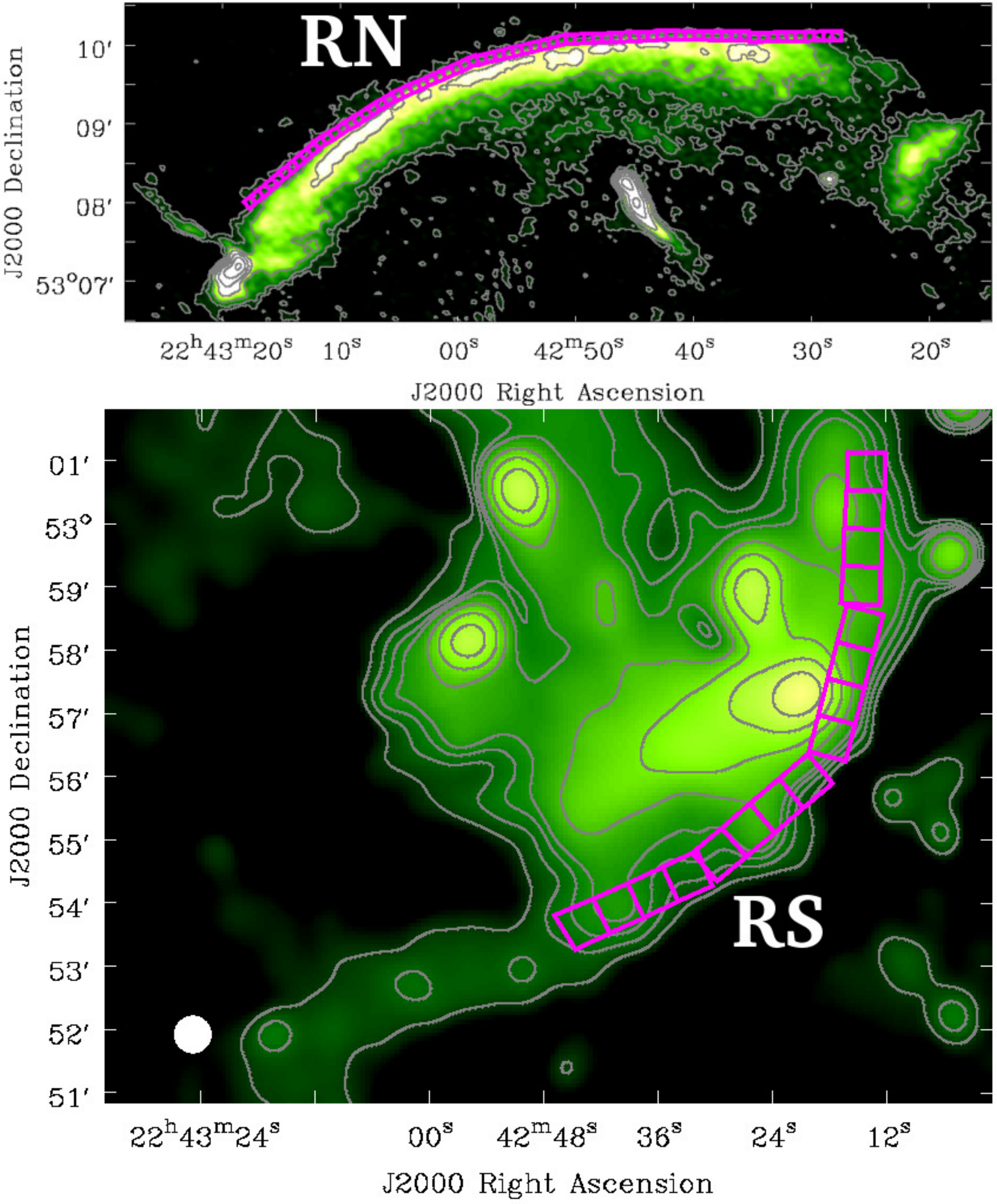}
\par\end{center}%
\end{minipage}\hfill{}%
\begin{minipage}[t]{0.53\textwidth}%
\begin{center}
\includegraphics[width=1\textwidth]{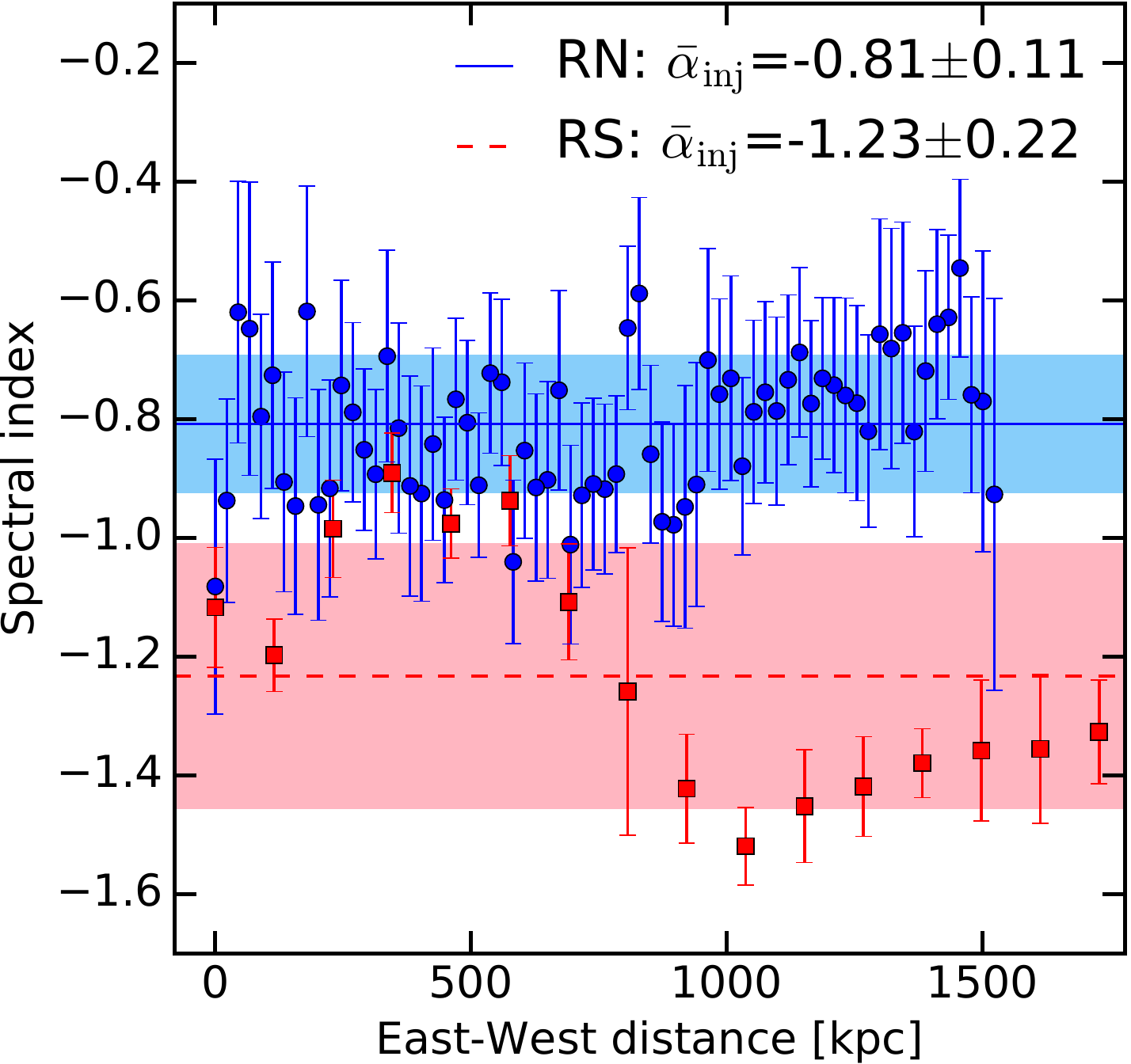}
\par\end{center}%
\end{minipage}\caption{Spectral index profiles (right) for the flat spectrum regions of RN
(top\textit{ }left) and RS (bottom\textit{ }left). The RN and RS profiles
were extracted from the 145-608 MHz spectral index map (Fig. \ref{fig:SPX_HLres})
and the 145 MHz - 2.3 GHz spectral index maps (Fig. \ref{fig:Halo_Spx_profile},
eight frequencies). The weighted means for RN ($\bar{\alpha}_{\text{inj}}^{n}=-0.81\pm0.11$)
and RS ($\bar{\alpha}_{\text{inj}}^{s}=-1.23\pm0.22$) are plotted
as horizontal lines (solid blue for RN and dashed red for RS). The
errors of the weighted means are shown by the filled regions of the
same colour. \label{fig:Spx_profile_along_RN_RS} }
\end{figure*}

From the $6.5\arcsec$-resolution spectral index map in Fig. \ref{fig:SPX_HLres},
 we found $\alpha_{\text{inj}}^{n}=-0.81\pm0.11$ within the shock
front regions in Fig. \ref{fig:Spx_profile_along_RN_RS}. This injection
index corresponds to $\mathcal{M}_{n}=2.7{}_{-0.3}^{+0.6}$, which
is consistent with the X-ray and spectral age modelling derived Mach
numbers (e.g. $2.90_{-0.13}^{+0.10}$ in \citealt{Stroe2014g} and
$2.7_{-0.4}^{+0.7}$ in \citealt{Akamatsu2015}). A low Mach number
like this may also be expected from structure formation simulations
which typically have $2\apprle\mathcal{M}\apprle4$ for internal shocks
(\citealt{Ryu2003,Pfrommer2006,Vazza2009}). For RS, since the shock
front location is detected furtherest from the cluster centre  in
the $35\arcsec$-resolution image (Fig. \ref{fig:Halo_Spx_profile}),
we estimated the injection index from the lower-resolution map (see
Fig. \ref{fig:Spx_profile_along_RN_RS}). We obtained an average value
of the injection index of $\alpha_{\text{inj}}^{s}=-1.23\pm0.22$
for RS. This is equivalent to a shock with Mach number of $\mathcal{M}_{s}=1.9_{-0.2}^{+0.3}$.
Similarly to RN, the Mach number for RS that is measured directly
at the shock front location in our resolved spectral index maps is
in agreement with those measured with X-ray data whereas the Mach
number derived from the integrated spectral index is $\sim2\sigma$
higher. (e.g. $\mathcal{M}_{s}^{X}=1.7_{-0.3}^{+0.4}$ in \citealt{Akamatsu2015},
see table \ref{tab:Spx_Mach}). It is noticed in Fig. \ref{fig:Spx_profile_along_RN_RS}
that the spectral indices along the shock front of RS have large variations
from a minimum of $-1.53\pm0.08$ (at $1.04\,\text{Mpc}$) to a maximum
of $-0.91\pm0.11$ (at $345\,\text{kpc}$), whereas the spectral indices
at the shock front in RN remain approximately constant. Additionally,
the spectral indices in the eastern side of RS (mean value of $\bar{\alpha}=-1.04\pm0.12$)
are flatter than those in the western side ($\bar{\alpha}=-1.43\pm0.08$).
 This could imply that the southern shock propagated with different
Mach numbers in different areas  and that the shock travelled with
a higher Mach number in the east. Another possibility is that due
to the complex morphology of RS the regions, where the spectral indices
were extracted, can host electron populations of different spectra
along the line of sight.

\subsubsection{Measurement uncertainty}

The estimation of the injection indices from the volume-integrated
and spatially-resolved spectra both have pros and cons. Firstly, while
the accuracy of the former method does not depend on the spatial resolution
of the observations nor the orientation of the merger axis, it may
be affected by other sources, since the radio-integrated fluxes are
calculated within the large volume. This argument was invoked to explain
Mach number discrepancies in several other cases (e.g. \citealt{ogrean2014a,Trasatti2015,Itahana2015}).
It is also very likely a contributing factor in the measurements of
RS where contaminating radio sources are present. Similarly, for RN
the integrated fluxes might be also contaminated in this case by additional
compression in the post-shock region that we discussed in Subsec.
\ref{subsec:RN}. To quantify this possible contamination, we estimated
the integrated spectrum of RN masking out the affected compression
region (Subsec. \ref{subsec:RN}). We obtained $\alpha_{\text{int}}^{n}=-1.08\pm0.04$
and $\mathcal{M}_{n}=5.1_{-0.9}^{+2.0}$ which is a higher Mach number
than we obtained prior to removing the potentially contaminated region.
Therefore, this potential contamination cannot explain the Mach number
discrepancy. Secondly, as mentioned earlier, the $\alpha_{\text{inj}}-\alpha_{\text{int}}$
relation (Eq. \ref{eq:Inj_Int_Approx}) does not hold for spherically
expanding shock waves such as RN and RS (\citealt{Kang2015a,Kang2015b}).
Therefore, the volume-integrated spectra will not necessarily characterise
the large-scale spherical shocks. This reason was used to explain
the discrepancy of the radio and X-ray derived Mach numbers in the
Toothbrush cluster (\citealt{VanWeeren2016b}). Thirdly, the discrepancy
might also come from micro-physics of the dynamics of relativistic
particles, including diffusion across magnetic filaments, re-acceleration
due to the interaction of CRs with magnetic field perturbations, adiabatic
expansion and changes of the magnetic field strength in the downstream
region (e.g. \citealt{Donnert2016}).

\begin{figure}
\centering{}%
\noindent\begin{minipage}[t]{1\columnwidth}%
\begin{center}
\includegraphics[clip,width=1\columnwidth]{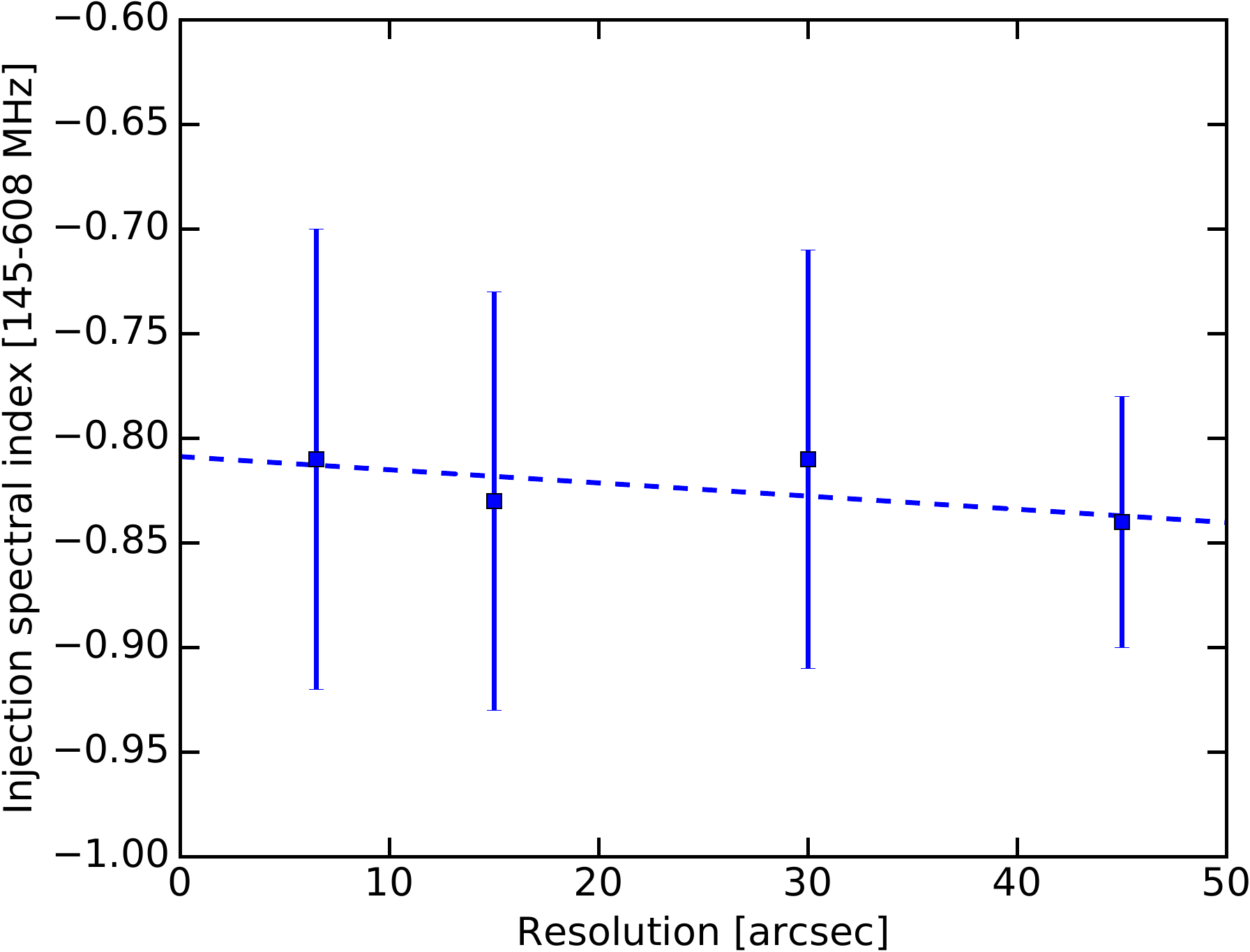}
\par\end{center}%
\end{minipage}\caption{Injection spectral indices between 145 and 608 MHz of RN as measured
at various spatial resolutions ($\alpha_{\text{inj}}=-0.81\pm0.63\times10^{-3}\theta_{[\text{arcsec}]}$).
The regions where the indices were extracted are similar to those
in Fig. \ref{fig:Spx_profile_along_RN_RS}.\label{fig:Spx_inj_vs_resolutions}
}
\end{figure}

The injection indices that are directly estimated from spatially resolved
spectral index maps might be less susceptible to contamination from
embedded sources. However, for such measurements to be accurate it
requires (\textit{i}) the merger axis to be on/close to the plane
of the sky to avoid the mixture of electron populations of different
velocities along the line of sight, (\textit{ii}) highly-resolved
spectral index maps to minimise the convolution effect caused by the
synthesised beam as this would artificially bias the injection indices,
and (\textit{iii}) precise alignment among the radio maps. In the
case of CIZA2242, hydrodynamical binary-merger simulations by \citet{VanWeeren2011}
suggested that the merger axis is likely to be close to the plane
of the sky (less than $10\degree$), and this was supported by \citet{Kang2012}
who performed DSA simulations. Moreover, the surface brightness along
the width of the eastern side of RN (Fig. \ref{fig:Post-Shock_Profiles},
red curve) drops  to $10\%$ of the peak value at a distance of $\sim130\,\text{kpc}$
from the peak which is consistent with the simulated profiles with
a   small viewing angle in\textbf{ }\citet{VanWeeren2011}. This implies
that RN satisfies (\textit{i}). However, due to the complex morphology
of RS, the southern shock might be observed at a larger angle than
the northern shock. To assess the impact of (\textit{ii}), we calculated
the injection indices at various spatial resolutions for RN. We made
spectral index maps between $145$ and $608\,\text{MHz}$ at resolutions
$6.5\arcsec,\,15\arcsec,\,30\arcsec$ and $45\arcsec$. The injection
indices were  measured at the shock-front  and the sizes of the square
regions used to extract the injection index are approximately equal
to the beam size of the maps (see Fig. \ref{fig:Spx_profile_along_RN_RS}).
We found that although the apparent injection locations are slightly
shifted to the North for the lower-resolution maps, the values of
the injection indices are weakly affected by the spatial resolution
(see Fig. \ref{fig:Spx_inj_vs_resolutions}). Assuming a linear relation
between $\alpha_{\text{inj}}$ and resolution $\theta$, we found
$\alpha_{\text{inj}}^{n}=(-0.81\pm0.10)+(-0.63\pm2.78)\times10^{-3}\theta[\text{arcsec}]$,
which is consistent with the value of $\alpha_{\text{inj}}^{n}=-0.81\pm0.11$
that we found in the $6.5\arcsec$-resolution map (Fig. \ref{fig:Spx_profile_along_RN_RS}).
To account for (\textit{iii}), we aligned the radio images when making
spectral index maps using the  procedure described in Subsec. \ref{subsec:Spectral_Index_maps}.
The GMRT $608\,\text{MHz}$ image was shifted a distance of $0.20\arcsec$
and $0.01\arcsec$ in RA and Dec axes, respectively. We found a small
increase of $1\%$ in the Mach number when aligning the images.

A potential further complication is that X-ray derived Mach numbers
also suffer from several systematic errors. As discussed in \citet{Akamatsu2017}
(see Sect.4.3 of their paper), there are three main systematics, which
prevents proper Mach number estimation from X-ray observations: (\textit{i})
the projection effect, (\textit{ii}) inhomogeneities in the ICM, and
(\textit{iii}) ion-electron non-equilibrium situation after the shock
heating. In case of CIZA2242, the first point was already discussed
above. Related to point (\textit{ii}), \citet{VanWeeren2011} revealed
that it is less likely to be a large clumping factor because of the
smooth shape of the relic. The third point would lead to the underestimation
of the Mach numbers from X-ray observations. However, it is difficult
to investigate this systematic without better spectra than those from
the current X-ray spectrometer. The upcoming Athena satellite can
shed new light on this point.

\begin{table*}
 \caption{Spectral indices and Mach numbers for the shock waves. \label{tab:Spx_Mach}}
 \centering{}
 \begin{tabular}{ccccccc}
 \hline
Source & $\alpha_{\text{inj}}$ & $\alpha_{\text{int}}$ & $\mathcal{M}_{\textrm{inj}}$ & $\mathcal{M}_{\textrm{int}}^{e}$ & $\mathcal{M}^{X}$ & $\mathcal{M}_{\textrm{ref}}$\tabularnewline
 \hline 
RN & $-0.81\pm0.11$ & $-1.11\pm0.04$ & $2.7_{-0.3}^{+0.6}$ & $4.4_{-0.6}^{+1.1}$ & $2.7_{-0.4}^{+0.7}$ $^{a}$ & $4.6_{-0.9}^{+1.3}$ $^{b}$, $4.58_{-1.09}^{+1.09}$ $^{c}$, $2.90_{-0.13}^{+0.10}$
 $^{d}$\tabularnewline
RS & $-1.23\pm0.22$ & $-1.41\pm0.05$ & $1.9_{-0.2}^{+0.3}$ & $2.4_{-0.1}^{+0.1}$ & $1.7_{-0.3}^{+0.4}$ $^{a}$ & $2.80_{-0.19}^{+0.19}$ $^{c}$\tabularnewline
R1 & $-0.91\pm0.14$ & $-1.18\pm0.06$ & $2.4_{-0.3}^{+0.5}$ & $3.5_{-0.4}^{+0.7}$ & $2.5_{-0.2}^{+0.6}$ & N/A\tabularnewline
 \hline
 \end{tabular}\\
$^{a}$: \citet{Akamatsu2015}, $^{b}$: \citet{VanWeeren2010a},
$^{c}$: \citet{Stroe2013a}, $^{d}$: \citet{Stroe2014g}, $^{e}$:
using Eq. \ref{eq:Mach} and \ref{eq:Inj_Int_Approx}. 
\end{table*}

\subsection{Particle acceleration efficiency}

\begin{figure}
\centering{}\includegraphics[width=1\columnwidth]{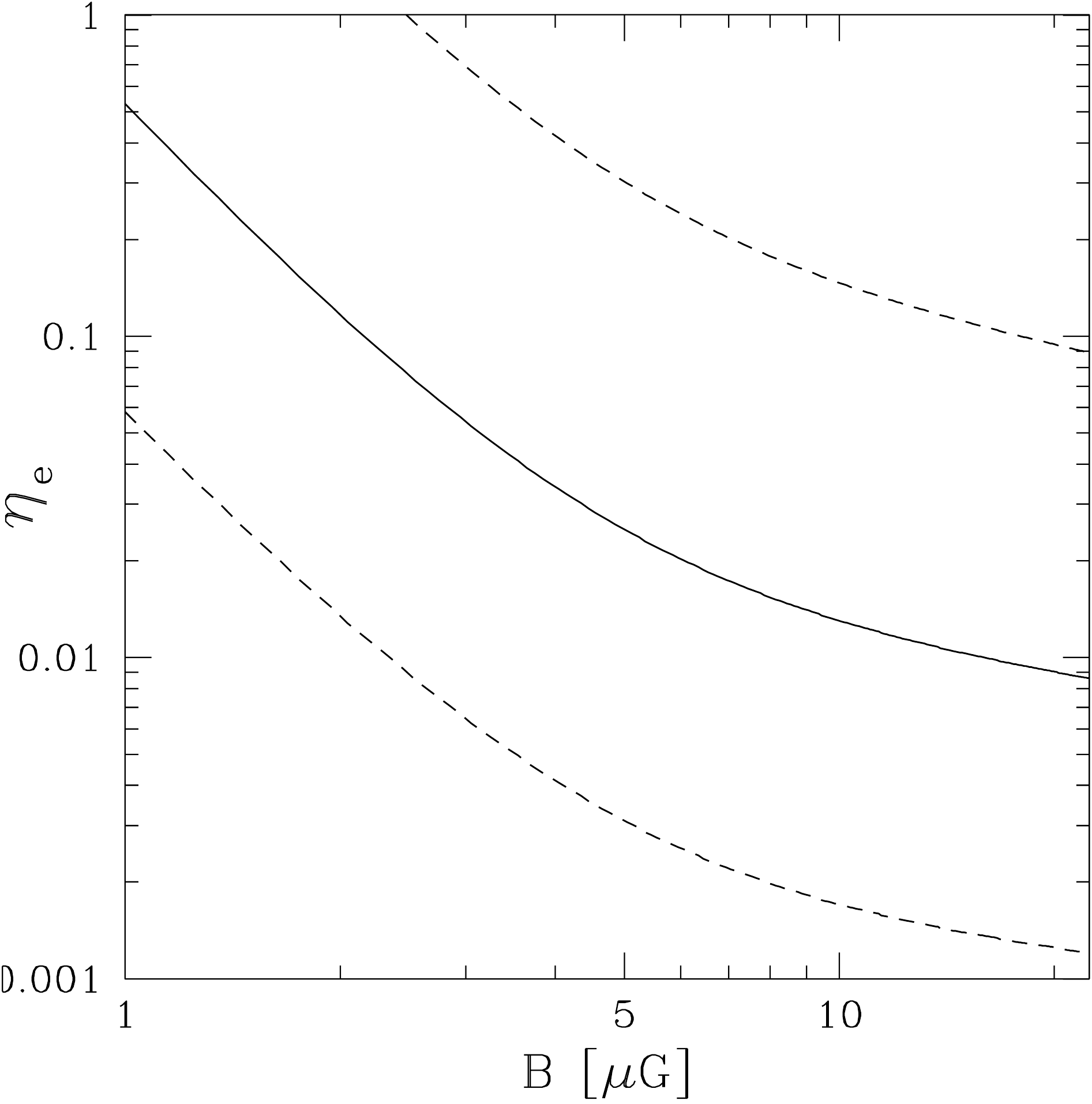}\caption{Particle acceleration efficiency as a function of magnetic field for
the northern relic. The solid curve is estimated for the Mach number
of $\mathcal{M}_{n}=2.7$ ($\alpha_{\text{inj}}=-0.81$). The dashed
curves are the $1\sigma$ uncertainties; the upper and lower dashed
curves are for the $\mathcal{M}_{n}=2.4$ ($\alpha_{\text{inj}}=-0.92$)
and $\mathcal{M}_{n}=3.3$ ($\alpha_{\text{inj}}=-0.70$) limits,
respectively. \label{fig:Efficiency}}
\end{figure}

The northern relic is known as one of the most luminous radio relics
so far. But the low Mach number of $\mathcal{M}_{n}=2.7_{-0.3}^{+0.6}$
for RN (Subsec. \ref{subsec:Mach_Number}) leads us to a question
of acceleration efficiency that is required to explain such a luminous
relic via shock acceleration of thermal electrons. The acceleration
efficiency was defined as the fraction of the kinetic energy flux
available at the shock that is converted into the supra-thermal and
relativistic electrons (\citealt{VanWeeren2016b}),
\begin{equation}
\eta_{e}=\frac{\epsilon_{e,\,\text{down}}v_{\text{down}}}{\Delta F_{\text{KE}}},
\end{equation}
where $\epsilon_{e,\,\text{down}}$ and $v_{\text{down}}$ are the
energy density and the velocity of the downstream accelerated electrons,
respectively; $\Delta F_{\text{KE}}=0.5\rho_{\text{up}}v_{s,\,\text{up}}^{3}\left(1-\frac{1}{C^{2}}\right)$
is the kinetic energy available at the shock; here $\rho_{\text{up}}$
and $v_{s,\,\text{up}}$ are the upstream density and the shock speed,
respectively; $C=\nicefrac{\mathcal{M}^{2}(\gamma+1)}{[2+(\gamma-1)\mathcal{M}^{2}]}$,
here $\gamma$ is the adiabatic index of the gas and $\mathcal{M}$
is the shock Mach number. Following formula in \citet{Botteon2016}\footnote{note that the left-hand side of Eq. 5 in \citet{Botteon2016} should
be $1/\Psi$.}, in Fig. \ref{fig:Efficiency} we report the efficiency of particle
acceleration by the northern shock as a function of the magnetic field
for the Mach number of $\mathcal{M}_{n}=2.7_{-0.3}^{+0.6}$ ($\alpha_{\text{inj}}=-0.81\pm0.11$).
We plotted the $1\sigma$ curves where $\mathcal{M}_{n}=2.4$ ($\alpha_{\text{inj}}=-0.92$,
upper dashed) and $\mathcal{M}_{n}=3.3$ ($\alpha_{\text{inj}}=-0.70$,
lower dashed), corresponding to the $1\sigma$ lower and upper boundaries,
respectively. The magnetic field range ($\lesssim20\,\text{\ensuremath{\mu}G}$)
was selected to cover the values, $5-7\,\text{\ensuremath{\mu}G}$,
in \citet{VanWeeren2010a}. The relatively high efficiency of electron
acceleration that should be postulated to explain the relic challenges
the case of a shock with $\mathcal{M}_{n}<2.7$. In this case a population
of pre-existing relativistic electrons in the upstream region of the
shock should be assumed. However, in the DSA framework, it is still
possible to accelerate electrons to relativistic energies directly
from thermal pools in the case of $\mathcal{M}_{n}>2.7$. Future works
will provide more constraints on this point. 

\subsection{The radio halo \label{subsec:Halo_Bridge}}

\subsubsection{The spatial spectral variations \label{subsec:Halo_SPX_Variations}}

In Fig. \ref{fig:SS_Lres} we show our LOFAR detection of the large-scale
diffuse emission that connects RN and RS. This was previously reported
in the WSRT 1.4 GHz observations and was interpreted as a radio halo
by \citet{VanWeeren2010a}. However, its physical characteristics,
such as its flux and spectral index, have not been studied. With our
deep LOFAR image (Fig. \ref{fig:SS_Lres}), we estimated the halo
size within the $3\sigma_{\text{noise}}$ region to be $1.8\,\text{Mpc}\times830\,\mbox{kpc}$
and the integrated flux to be $S_{h}=346\pm64\,\textrm{mJy}$ (see
Subsec. \ref{subsec:RH}). The halo size in the north-south direction
was measured between the northern and southern relic inner edges.
The halo maintains its surface brightness, even in the regions that
are least contaminated by the tailed AGNs, such as the regions east
of source D or west of source E. The presence of the halo with elongated
morphology connecting the north and south relics suggests a connection
between the shock waves that are responsible for RN and RS. A comparable
example of the morphological connection between radio relic and halo
was observed in RX J0603.3+4214 where a giant radio relic at the northern
edge of the cluster is connected to an elongated uniform brightness
radio halo in the cluster centre (\citealt{VanWeeren2016b}).
\begin{center}
\begin{figure*}
\centering{}%
\begin{minipage}[t][1\totalheight][c]{0.33\textwidth}%
\begin{center}
\includegraphics[width=1\columnwidth]{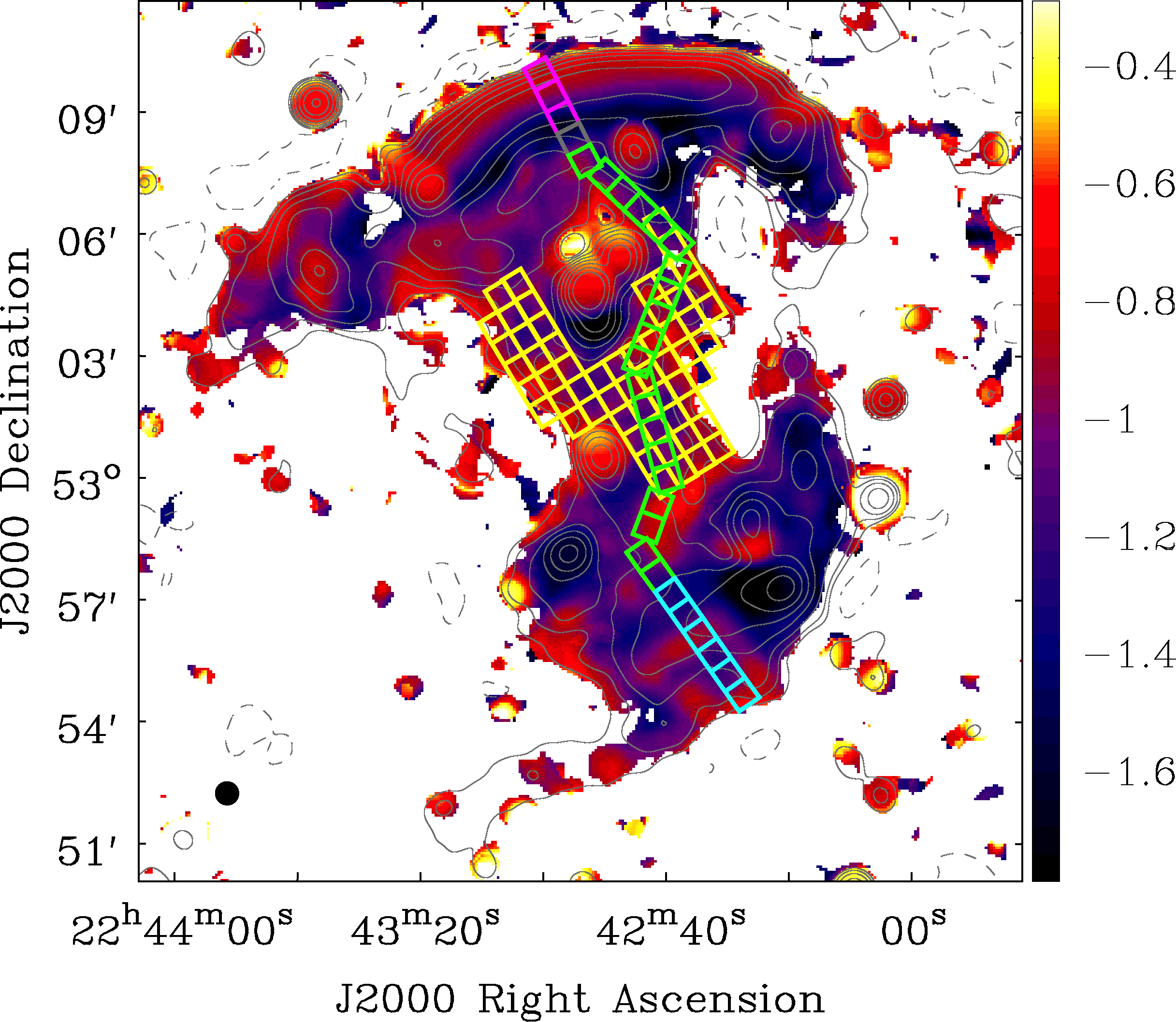}
\par\end{center}%
\end{minipage}\hfill{}%
\begin{minipage}[t][1\totalheight][c]{0.32\textwidth}%
\begin{center}
\includegraphics[width=0.9\columnwidth]{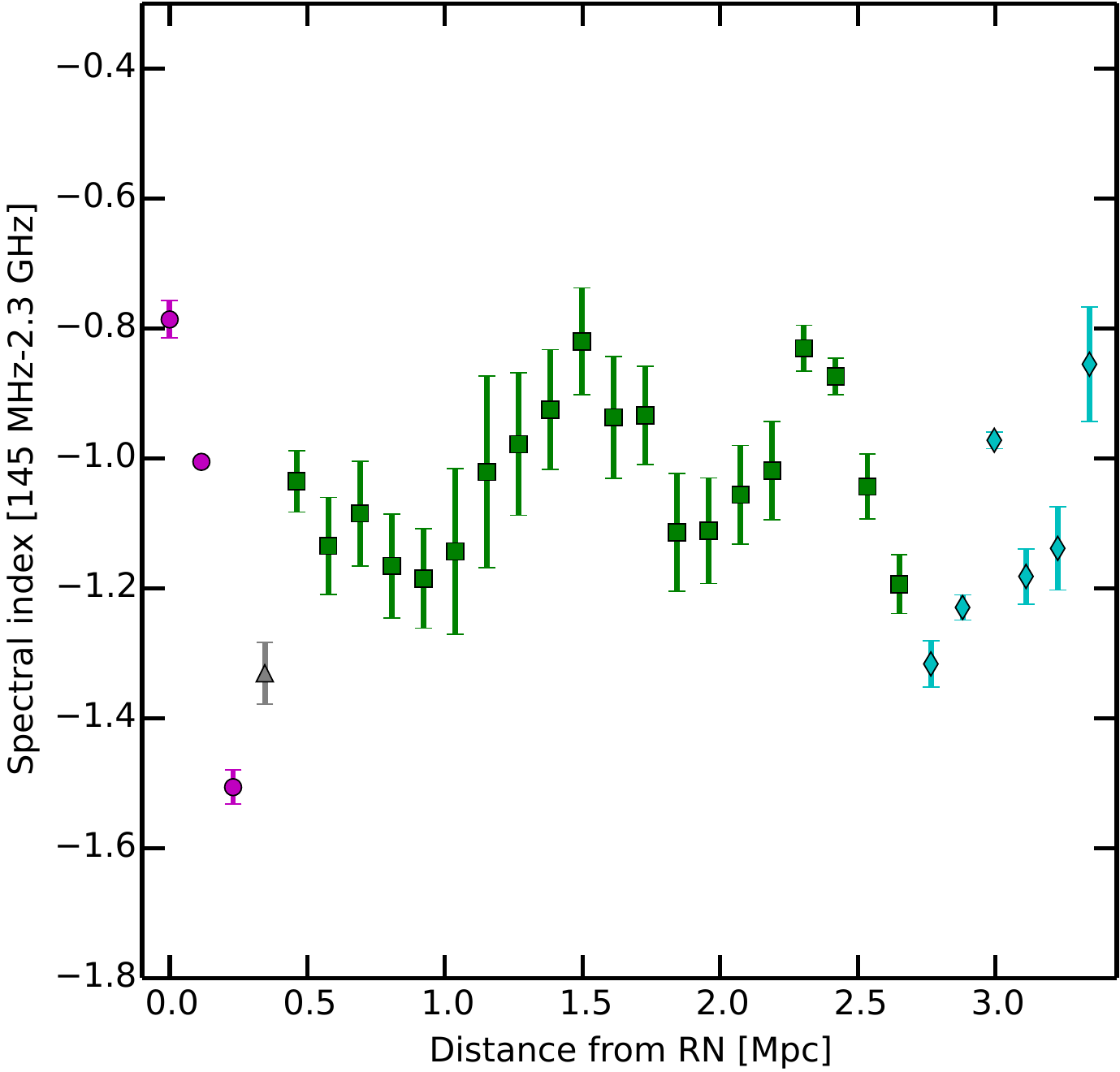}
\par\end{center}%
\end{minipage}\hfill{}%
\begin{minipage}[t][1\totalheight][c]{0.32\textwidth}%
\begin{center}
\includegraphics[width=0.9\columnwidth]{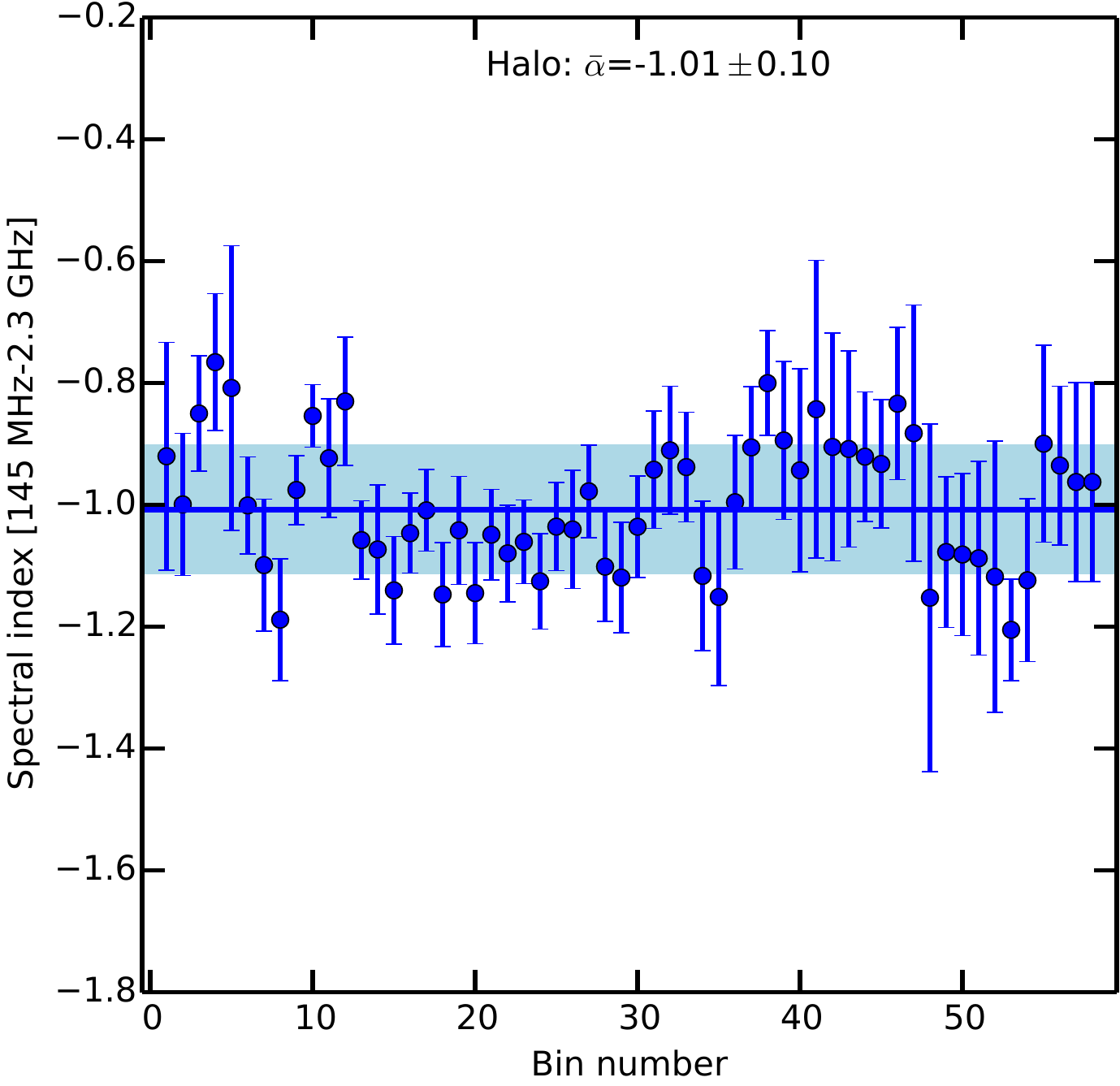}
\par\end{center}%
\end{minipage}\caption{Left: A $35\arcsec$-resolution spectral index map between 145 MHz
and 2.3 GHz (eight frequencies) including the regions where spectral
indices were extracted. The square regions have a width of is $36\arcsec$
($115\,\text{kpc}$). In the analysis of the spectral index variations, we did not take into account an absolute flux scale uncertainty of 10\% as done for the $6.5\arcsec$ and $18\arcsec\times16\arcsec$ maps in Fig. \ref{fig:SPX_HLres} and \ref{fig:SPX_RS_R1_R2}, respectively. Middle: Spectral index profile for the square
(magenta, gray, green, cyan) regions in the left image. Right: The
spectral indices in the square yellow regions over the selected radio
halo area where the contamination from radio galaxies is minimised.
The inverse-invariance weighted mean of the spectral indices for the
halo regions (yellow squares ) is $-1.01\pm0.10$. \label{fig:Halo_Spx_profile}
 }
\end{figure*}
\par\end{center}

We examined a spectral index profile (see Fig. \ref{fig:Halo_Spx_profile},
middle) across the north and south relics and  the radio halo. We
found that the spectral index steepens from $-0.79\pm0.09$ at the
northern shock front location to $-1.51\pm0.08$ in the post-shock
region ($230\,\text{kpc}$). It then flattens over a distance of $115\,\text{kpc}$ and remains approximately constant with a mean of $-1.03\pm0.12$ over a distance of $2.5\,\mbox{Mpc}$. This mean spectral index of the halo is $-1.01\pm0.10$ when measured in a $\sim\text{1 Mpc}^2$ region (i.e. yellow squares in Fig. \ref{fig:Halo_Spx_profile}, left) and is consistent with the integrated spectral index of $-1.03\pm0.09$
 that was estimated in Subsec. \ref{subsec:RH}. Similarly, the spectral index profile across RS in the south-north direction (i.e. right-left direction in Fig. \ref{fig:Halo_Spx_profile}, middle) steepens from
$-0.86\pm0.08$ at the southern shock front to $-1.32\pm0.05$ over
a distance of $575\,\text{kpc}$ from the southern shock front (cyan
diamonds at $\sim2.7\,\text{Mpc}$ in Fig. \ref{fig:Halo_Spx_profile},
middle). After this steepening, the spectral indices flatten to $-0.83\pm0.05$
over a distance of $460\,\text{kpc}$ (i.e. $2.2-2.7\,\text{Mpc}$
in Fig. \ref{fig:Halo_Spx_profile}, middle). However, it is possible
that the spectral indices across RS and in the southern halo region
are strongly effected by contaminating sources (i.e. K, J). Moreover,
the spectral index variations in this area are at similar levels to
those seen across the halo and therefore cannot be associated with
the southern shock with certainty.

The spectral re-flattening of the emission downstream from the northern
relic and into the northern part of the halo suggests particles have
undergone re-acceleration. It is currently thought that giant radio
halos trace turbulent regions in which particles are re-accelerated
(see \citealt{Brunetti2014} for review). The fact that in several
cases X-ray shocks coincide with edges of radio halos is suggestive
of the possibility that these shocks can be sources of turbulence
downstream (e.g. \citealt{Markevitch2010,Shimwell2014}). More specifically
shocks may inject compressive turbulence in the ICM and compressive
turbulence is in fact used to model turbulent re-acceleration of electrons
in radio halos in several papers (e.g. \citealt{Brunetti2007a}).
In the case of CIZA2242 the electrons might be accelerated (or re-accelerated)
at the northern shock front and lose energy downstream via radiative
losses. However on longer times the shock-generated turbulence might
decay  to smaller scales and re-accelerate  electrons inducing a flattening
of the synchrotron spectrum. A similar scenario was invoked to explain
the spatial behaviour of the spectral index in the post-shock region
of RX J0603.3+4214 and the hole in the radio halo of Abell 2034 (\citealt{VanWeeren2016b}
and \citealt{Shimwell2016a}). In RX J0603.3+4214, \citet{VanWeeren2016b}
also observed spectral steepening in the post-shock region from $-0.8$
at the shock front edge to $-2$ at the boundary of the relic and
halo, which was followed by an approximately constant spectral index
of $-1.06\pm0.06$ across the Mpc-scale halo region. This suggests
that in these, and potentially other clusters, we may be observing
similar particle (re-)acceleration and ageing scenario.

The spectral indices over the large region of the radio halo, where
the contaminating sources (i.e. B, C, D, E, F, G, K, I) are masked,
are plotted in Fig. \ref{fig:Halo_Spx_profile} (right). Following
\citet{Cassano2013a}, we estimated the raw scatter (see Eq. 5 in
\citealt{Cassano2013a}) from the inverse-variance weighted mean to
be $0.10$. From the weighted mean of the spectral index errors of
$0.10$ unit that propagates from the image noise, the intrinsic scatter
of $\sqrt{0.10^{2}-0.09^{2}}=0.04$ was obtained. The intrinsic scatter
is smaller than the typical spectral index errors of $\Delta\alpha\thickapprox0.13$
in the halo region (yellow squares in Fig. \ref{fig:Halo_Spx_profile},
left); hence, the measurements of the spectral indices in the halo
region are close to the detection limit of our data. More precise
spectral index measurements would therefore be required to study the
physics of turbulence in the halo.

\subsubsection{A comparison with radio-thermal correlations}

\begin{figure}
\centering{}\includegraphics[width=1\columnwidth]{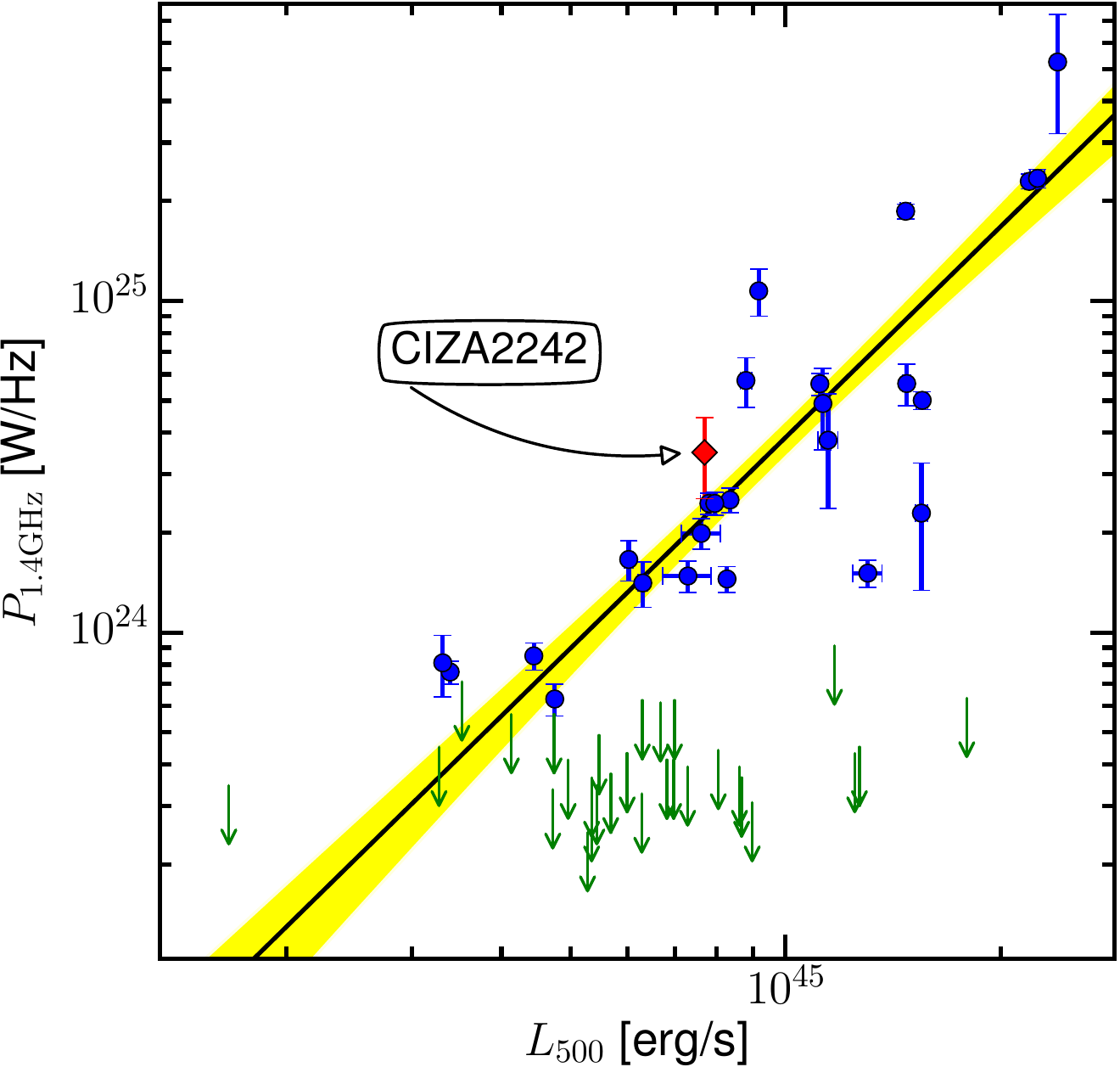}\caption{The scaling relation of the radio power $P_{1.4\,\text{GHz}}$ and
X-ray luminosity $L_{500}$  for radio halos including CIZA2242 (red
diamond). The list of the halos and the BCES bisector best fit $P_{1.4\,\text{GHz}}[10^{24}\,\text{W\,H\ensuremath{z^{-1}}}]=10^{-1.52\pm0.21}\times L_{500}^{2.11\pm0.20}[10^{44}\,\text{erg}\,\text{s}^{-1}]$
 (without CIZA2242) were given in \citet{Cassano2013a}. The shadowed
area is the $95\%$ confidence regions of the best-fit relation. \label{fig:Scaling_relation}}
\end{figure}

Galaxy clusters statistically branch into two populations (\citealt{Brunetti2007b,brunetti2009,Rossetti2011,Cassano2013a,Cuciti2015,Yuan2015}):
dynamically disturbed systems host radio halos whose luminosity correlates
with the X-ray luminosity and mass of the hosting systems, whereas
in general relaxed systems do not host Mpc-scale halos at the sensitivity
level of available observations. To examine whether or not the radio
halo of CIZA2242 follows the relationship between radio luminosity
and X-ray luminosity and mass of the hosting cluster, in Fig. \ref{fig:Scaling_relation}
 we plot the radio power $P_{1.4\,\text{GHz}}$ versus the X-ray luminosity
$L_{500}$ in the $0.1-2.4\,\text{keV}$ energy band for a number
of clusters that are given in \citet{Cassano2013a}; $L_{500}$ is
the luminosity within the radius $R_{500}$ where the ICM matter density
is 500 times the critical density of the Universe at redshift $z$.
 The radio power at 1.4 GHz for the CIZA2242 halo was estimated by
extrapolating our measurements  to be $P_{1.4\,\text{GHz}}=(3.5\pm1.0)\times10^{24}\,\text{W\,H}\text{z}^{-1}$
at the cluster redshift $z=0.192$ using the LOFAR 145 MHz integrated
flux of $346\pm64\,\text{mJy}$ and the integrated spectral index
of $-1.03\pm0.09$.  Using the Chandra data (\citealt{Ogrean2014}),
we measured the X-ray luminosity of $L_{500}=(7.7\pm0.1)\times10^{44}\,\text{erg\,\ensuremath{\text{s}^{-1}}}$
for the $0.1-2.4\,\text{keV}$ energy band within a radius of $R_{500}=1.2\,\text{Mpc}$
at $z=0.192$. Fig. \ref{fig:Scaling_relation} shows that the CIZA2242
data point  closely follows the $P_{1.4\,GHz}-L_{500}$ correlation
(i.e. $P_{1.4\,\text{GHz}}[10^{24}\,\text{W\,H\ensuremath{z^{-1}}}]=10^{-1.52\pm0.20}L_{500}{}^{2.11\pm0.20}[10^{44}\,\text{erg\,\ensuremath{\text{s}^{-1}}}]$;
BCES bisector best fit in \citealt{Cassano2013a}).  

Despite RN being amongst the most prominent radio relics known, the
radio halo of CIZA2242 seems  typical,  with moderate radio power
and X-ray luminosity in comparison with other radio halos (Fig. \ref{fig:Scaling_relation}).
Magneto-hydrodynamic simulations of re-acceleration of relativistic
electrons via post-shock generated turbulence by \citet{Donnert2013}
predict that the integrated spectrum of the halos is flatter in their
early phase ($\lesssim1\,\text{Gyr}$) and becomes steeper after $\sim1.2\,\text{Gyrs}$
of core passage. With our current data, the radio halo of CIZA2242 has a relatively flat
spectral index of $-1.03\pm0.09$ and is among the flatter spectrum
halos known (e.g. \citealt{Luigina2012}), implying that it has recently
formed (e.g. also see our discussion in Subsec. \ref{subsec:Halo_SPX_Variations}).
The early phase of the halo is further supported by binary cluster
merger simulations by \citet{VanWeeren2011} which suggested that
the cluster is $1\,\text{Gyr}$ after core passage.

\textbf{}

\subsection{A newly detected eastern shock wave? \label{subsec:New_shock}}

\begin{figure*}
\begin{minipage}[t]{0.49\textwidth}%
\begin{center}
\includegraphics[width=1\columnwidth]{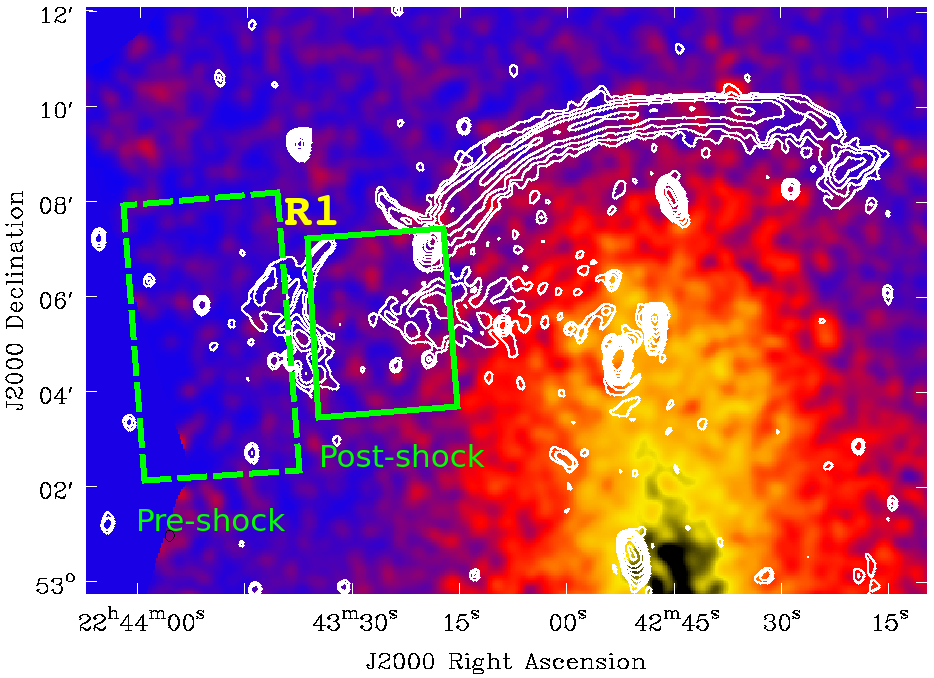}
\par\end{center}%
\end{minipage}%
\begin{minipage}[t]{0.45\textwidth}%
\begin{center}
\includegraphics[width=1\columnwidth]{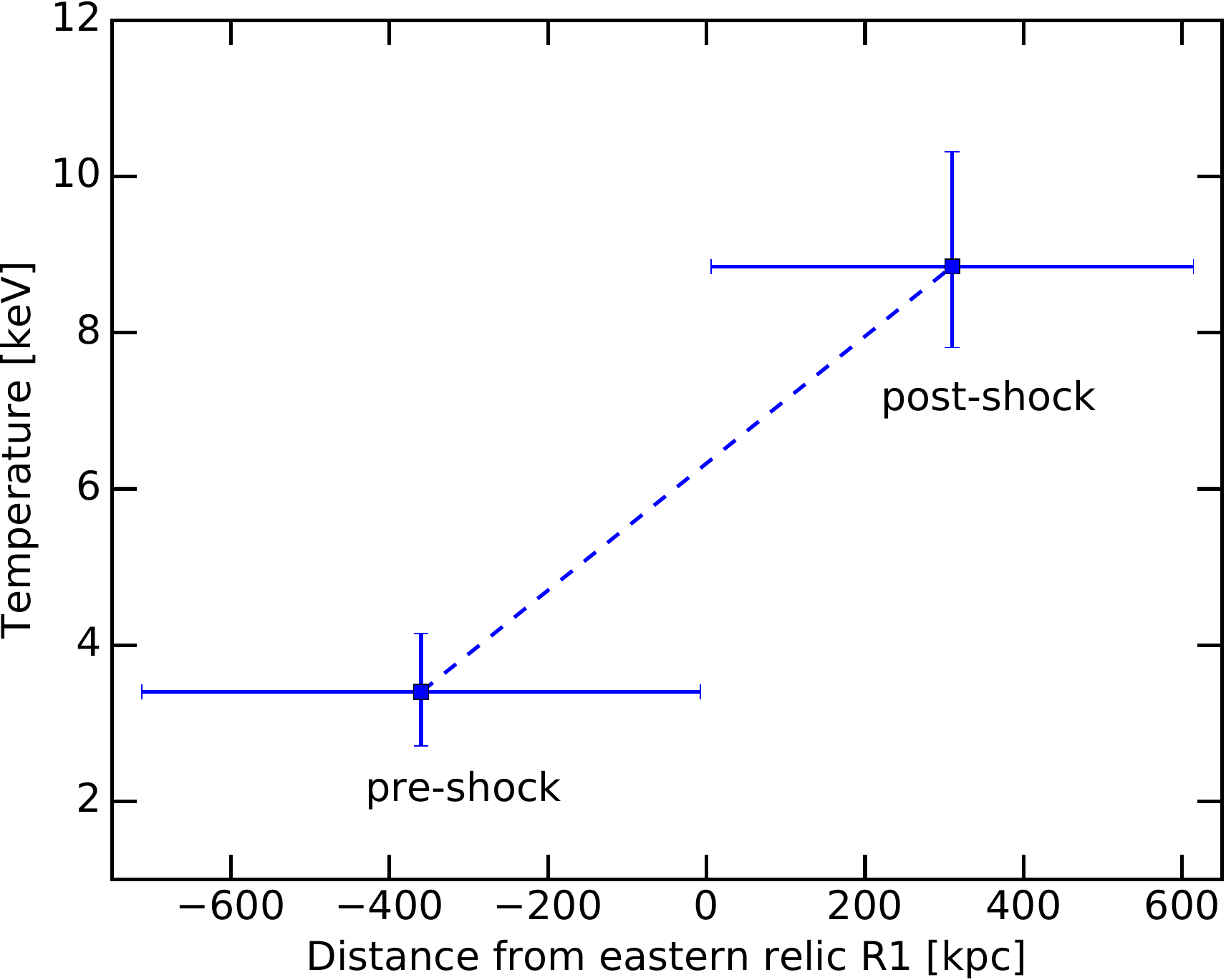}
\par\end{center}%
\end{minipage}\caption{Left\textit{: }Suzaku X-ray image for the north-east region of CIZA2242
overlaid with the WSRT 1.4 GHz contours (the first contour is $3\sigma_{\text{noise}}$,
$\sigma_{\text{noise}}=30\,\mbox{\ensuremath{\mu}Jy/beam}$, beam
size of $14\arcsec\times11\arcsec$, and the next ones are spaced
by 2.) The green rectangles are the pre- and post-shock regions of
R1 within which the X-ray temperatures were extracted. Right: X-ray
temperature for the corresponding regions in the left image.  \label{fig:Suzaku_regions}}
\end{figure*}

\begin{figure}
\centering{}\includegraphics[width=1\columnwidth]{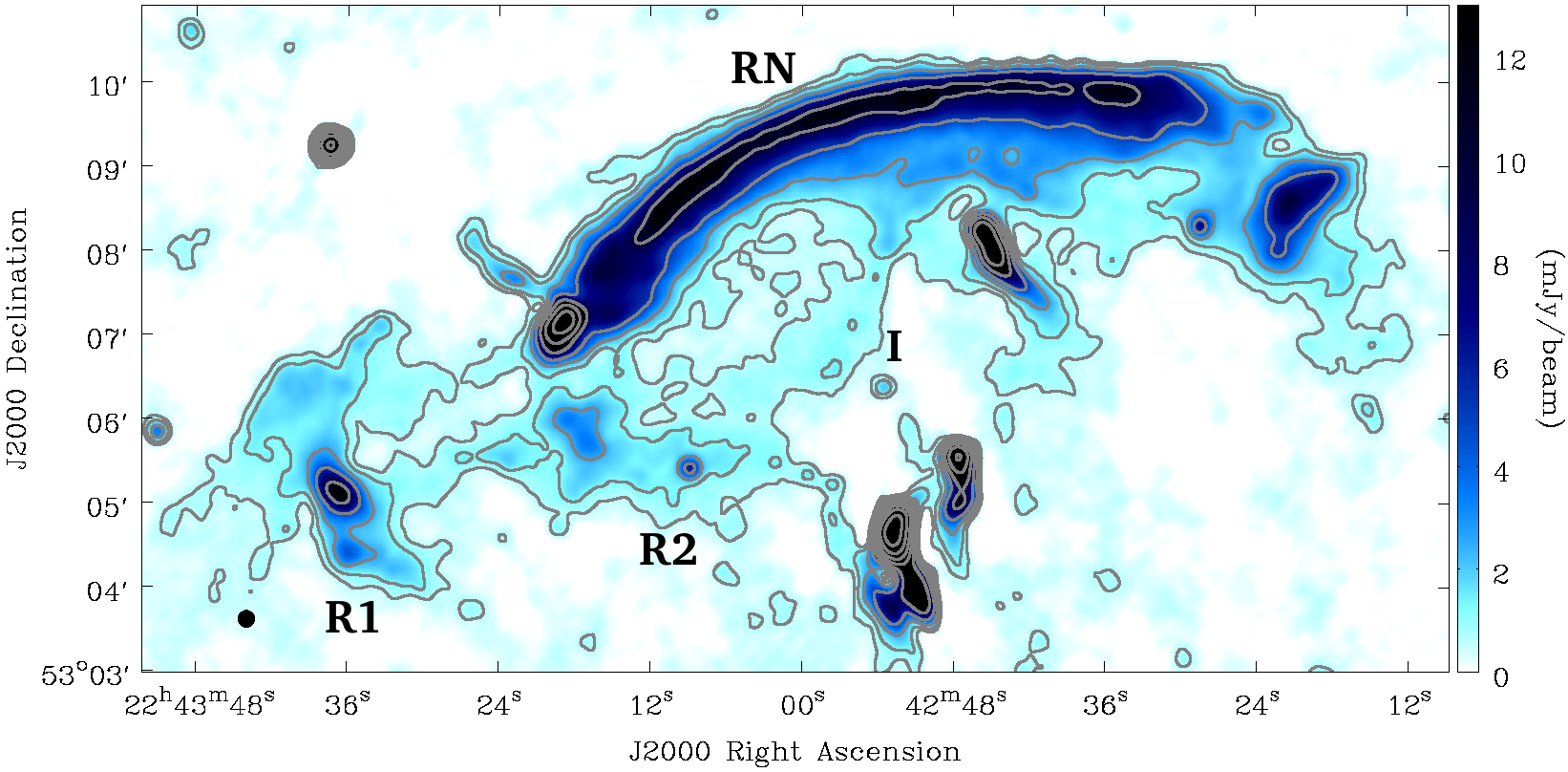}\caption{A zoom-in northern region of the LOFAR intensity image at $12\arcsec$
resolution. The first contour is at $3\sigma_{\text{noise}}$ ($\sigma_{\text{noise}}=210\,\text{\ensuremath{\mu}Jy/beam}$);
the next ones is spaced by 2. \label{fig:SS_Northern_region_12arcsec}
 }
\end{figure}

The steepening features in the spectral index of R1, reported in Subsec.
\ref{subsec:RE}, are likely due to the energy loses of relativistic
electrons (e.g. synchrotron, IC emission). Additionally, the intensity
maps (Fig. \ref{fig:SS_Hres}, \ref{fig:SPX_RS_R1_R2} and \ref{fig:Suzaku_regions},
left) show emission with a sharp edge on the eastern side and a more
gradual decline towards the  western edge. These features are among
the typical properties of radio emission associated with a shock wave.
We speculate that R1E and R1W trace a complex  shock wave that is
propagating away from the cluster centre and that this shock (re-)energises
two clouds of relic or aged electrons. \textbf{} In this scenario,
the spectral steepening is more visible for R1E which implies that
the shock wave is more on the plane of the sky in this region.. A
$12\arcsec$ resolution map in Fig. \ref{fig:SS_Northern_region_12arcsec}
shows a bridge of low surface brightness (a $3\sigma_{\text{noise}}$
detection) that connects R1, R2, I, and RN. This  suggests the origin
of the eastern shock may be closely related to the cluster major merger.
On our $18\arcsec\times16\arcsec$ resolution map (Fig. \ref{fig:SPX_RS_R1_R2},
right) we measured the injection indices of $-0.89\pm0.08$ for R1E
and $-0.92\pm0.12$ for R1W, the  average of these injection indices
is $-0.91\pm0.14$ which is equivalent to a Mach number of $2.4_{-0.3}^{+0.5}$.
Our estimation of the integrated spectral index of R1, within the
LOFAR $>3\sigma_{\text{noise}}$ region, masking out the compact source
in the central of R1W, is $-1.18\pm0.06$. This is equivalent to a
shock Mach number of $3.5_{-0.4}^{+0.7}$. Similarly to RN and RS,
this Mach number is higher than the value that was calculated from
the high resolution spectral index map.

To search for imprints of a shock in the X-ray data, we re-analysed
the Suzaku data (\citealt{Akamatsu2015}) in the pre- and post-shock
regions of R1 (see Fig. \ref{fig:Suzaku_regions}). We found an average
temperature jump from $kT_{\text{pre}}=3.5{}_{-0.5}^{+0.8}\,\mbox{keV}$
to $kT_{\text{post}}=9.6{}_{-1.1}^{+1.5}\,\mbox{keV}$ from the pre-
to post-shock region. This temperature jump of $6.1\,\text{keV}$
across the eastern relic is 2.4 times higher than a temperature decrease
of $\sim2.5\,\text{keV}$ over the eastern, slightly south region
(i.e. centred at RA=22:43:24, DEC=+52.57.00) that is at a similar
radial distance as the eastern relic but is devoid of radio emission
(see Fig. 6 and 7 in \citealt{Akamatsu2015}). Therefore, the high
temperature jump ($6.1\,\text{keV}$) is likely associated with the
eastern relic. This X-ray temperature jump corresponds to a Mach number
of $2.5_{-0.2}^{+0.6}$, which is consistent with our radio derived
Mach numbers. The X-ray detection of a shock wave could be confirmed
by measuring the discontinuity of the X-ray surface brightness profile.
However, the existing Chandra and XMM-Newton data (\citealt{Ogrean2013a,Ogrean2014,Akamatsu2015})
do not have sufficient spatial coverage to detect the surface brightness
discontinuity at the location of the eastern relic; and the Suzaku
data has a limited spatial resolution of $2\arcmin$ which is insufficient
for the surface brightness analysis.\textbf{ } Our collection of radio
and X-ray measurements provides compelling evidence that R1 is tracing
another shock front in CIZA2242 that is propagating eastwards.

\textbf{}

\section{Conclusions}
\label{sec:Conclusions}
 
We have presented deep, high-fidelity LOFAR 145MHz images of CIZA2242,
which have a resolution of $\sim5\arcsec$ and a sensitivity of $140\,\text{\ensuremath{\mu}Jy/beam}$.
The LOFAR data, in combination with the existing GMRT, WSRT, Suzaku
and Chandra data were used to create spectral index maps of CIZA2242
at resolutions of $6.5\arcsec$, $18\arcsec\times16\arcsec$ and $35\arcsec$.
Below we summarise our main results.

\begin{itemize}
\item To investigate the long-standing discrepancy between X-ray and radio
derived Mach numbers, we have measured injection spectral indices
of $-0.81\pm0.11$ and $-1.23\pm0.22$ for the northern and southern
relics, respectively. These correspond to Mach numbers of $\mathcal{M}_{n}=2.7_{-0.3}^{+0.6}$
and $\mathcal{M}_{s}=1.9_{-0.2}^{+0.3}$, and are in agreement with
the Mach numbers derived from X-ray data (e.g. $\mathcal{M}_{n}=2.7_{-0.4}^{+0.7}$
and $\mathcal{M}_{s}=1.7_{-0.3}^{+0.4}$ in \citealt{Akamatsu2015})
and spectral age modelling study of the radio emission (e.g. $\mathcal{M}_{n}=2.9_{-0.13}^{+0.10}$
in \citealt{Stroe2014g}). 
\item We have confirmed the existence of a radio halo and constrained its
integrated flux ($346\pm64\,\text{mJy}$) and its integrated spectral
index ($-1.03\pm0.09$, between 145 MHz and 2.3 GHz). We measured
the radio halo power $P_{1.4\,\text{GHz}}=(3.5\pm1.0)\times10^{24}\,\text{W\,H}\text{z}^{-1}$
and the X-ray luminosity $L_{500}=(7.7\pm0.1)\times10^{44}\text{erg}\,\text{s}^{-1}$,
which is close to  the value expected from the $P_{1.4\,\text{GHz}}$-$L_{500}$
correlation. We have discussed a possible connection between the northern
and southern relics and the halo, and have speculated that the formation
of the halo may be driven by turbulence generated by the passing shock
waves.
\item In the radio source R1, in the north eastern region of the cluster,
we found spectral steepening towards the cluster centre. A temperature
jump from $3.5_{-0.5}^{+0.8}\,\text{keV}$ to $9.6_{-1.1}^{+1.5}\,\text{keV}$
was also detected at the location of the eastern relic R1 by re-analysing
the existing Suzaku X-ray data. We suggest that R1 is an eastern relic
that traces a shock wave that is propagating eastwards and (re-)energises
the ICM electrons. We estimated an injection spectral index of $-0.91\pm0.14$
and a Mach number of $\mathcal{M}_{e}=2.4_{-0.3}^{+0.5}$, which is
consistent with our re-analysis of the Suzaku data from which we derived
$\mathcal{M}_{e}^{X}=2.5_{-0.2}^{+0.6}$.\textbf{}
\end{itemize}

\section*{Acknowledgements}

We thank the anonymous referee for the helpful comments. This paper
is based (in part) on results obtained with LOFAR equipment. LOFAR
(\citealt{VanHaarlem2013}) is the Low Frequency Array designed and
constructed by ASTRON. We thank the staff of the GMRT who have made
these observations possible. The GMRT is run by the National Centre
for Radio Astrophysics of the Tata Institute of Fundamental Research.
The Westerbork Synthesis Radio Telescope is operated by the ASTRON
(Netherlands Institute for Radio Astronomy) with support from the
Netherlands Foundation for Scientific Research (NWO). The scientific
results reported in this article are based in part on observations
made by the Chandra X-ray Observatory and published previously in
\citet{Ogrean2014}. This research has made use of data obtained from
the Suzaku satellite, a collaborative mission between the space agencies
of Japan (JAXA) and the USA (NASA).  DNH, TS and HR acknowledge support
from the ERC Advanced Investigator programme NewClusters 321271. DDM
acknowledges support from ERCStG 307215 (LODESTONE). GB and RC acknowledge
partial support from PRIN INAF 2014 and JD acknowledges support from
ERC Marie-Curie Grant 658912 (Cosmo Plasmas). GJW gratefully acknowledges
support from The Leverhulme Trust. HA acknowledges the support of
NWO via a Veni grant. SRON is supported financially by NWO, the Netherlands
Organization for Scientific Research. MB and MH acknowledge support
by DFG FOR 1254. AD acknowledges support by BMBF 05A15STA. We thank
G. A. Ogrean for providing us the Chandra map of CIZA2242. We thank
M. James Jee for discussion on the weak lensing mass of CIZA2242. 

\bibliographystyle{mnras}
\bibliography{pp_SS_LOFAR}

\appendix

\section{Integrated fluxes for the radio relics and halo}
\label{sec:Integrated_fluxes}

Table \ref{tab:Int_fluxes} shows the integrated fluxes for the radio relics and halo in CIZA2242 that are plotted in Fig. \ref{fig:Spx_Int_RN_RS_Halo}.  Note that the integrated fluxes for RN were reported in \citet{Stroe2015d}; here we present the integrated fluxes from the images that were made with different CLEANing parameters (see Table \ref{tab:Imaging_Parameters}).

\begin{table}
	
	\caption{Integrated fluxes for the radio relics and halo in CIZA2242 in Fig. \ref{fig:Spx_Int_RN_RS_Halo}. The integrated fluxes for the relics were measured from the $18\arcsec\times16\arcsec$ images, and the halo fluxes were estimated  from the $35\arcsec$ images (see Table \ref{tab:Imaging_Parameters} for imaging parameters). The flux measurement errors for the relics were added absolute flux scale uncertainties of $10\%$ of the integrated fluxes. The flux error estimation for the radio halo was described in Subsec. \ref{subsec:RH}.}
	
	\begin{tabular}{cccccc}
		 \hline 
		Freq. & RN & RS & Halo & R1 & R2 \tabularnewline
		(MHz) &(mJy) &(mJy)&(mJy)&(mJy)&(mJy)\tabularnewline
		 \hline 
		$145$ & $1637\pm168$ & $777\pm82$ & $346\pm64$ & $144\pm17$ & $143\pm16$\tabularnewline
		$153$ & $1488\pm171$ & $711\pm93$ & $288\pm64$ & $76\pm20$ & $122\pm24$\tabularnewline
		$323$ & $646\pm71$ & $193\pm25$ & --$^a$ & $38\pm7$ & $20\pm5$\tabularnewline
		$608$ & $337\pm35$ & $83\pm9$ & $59\pm20$ & $16\pm2$ & $29\pm3$\tabularnewline
		$1221$ & $148\pm16$ & $30\pm4$ & $28\pm10$ & $13\pm2$ & $16\pm2$\tabularnewline
		$1382$ & $140\pm14$ & $34\pm4$ & $43\pm10$ & $10\pm1$ & $15\pm2$\tabularnewline
		$1714$ & $106\pm11$ & $22\pm3$ & $27\pm8$ & $6\pm1$ & $8\pm1$\tabularnewline
		$2272$ & $72\pm8$ & $17\pm2$ & $19\pm6$ & $4\pm1$ & $8\pm1$\tabularnewline
		 \hline 
	\end{tabular}\\
	$^{a}$: a flux of 135 mJy was measured in the LOFAR  $\geqslant3\sigma_{\text{noise}}$ region of the GMRT 323 MHz $35\arcsec$ map. However, most of the flux comes from the residuals of the subtracted compact sources. We excluded the halo flux at the 323 MHz in the integrated spectral index estimation.
	\label{tab:Int_fluxes}
\end{table}

\section{Spectral index error maps}
\label{sec:Spectral_Index_Error}

In Fig. \ref{fig:SPX_Error_Map_Hres}-\ref{fig:SPX_Error_Map_R1R2},
we show the error maps for the corresponding spectral index maps in
Fig. \ref{fig:SPX_HLres}, \ref{fig:SPX_RS_R1_R2} and \ref{fig:Halo_Spx_profile}. The error
estimation takes into account the individual image noise and a flux
scale error of $10\%$, which is formulated in Eq. \ref{eq:Spec_Index_Error}.

\begin{center}
\begin{figure}
\centering{}\includegraphics[width=1\columnwidth]{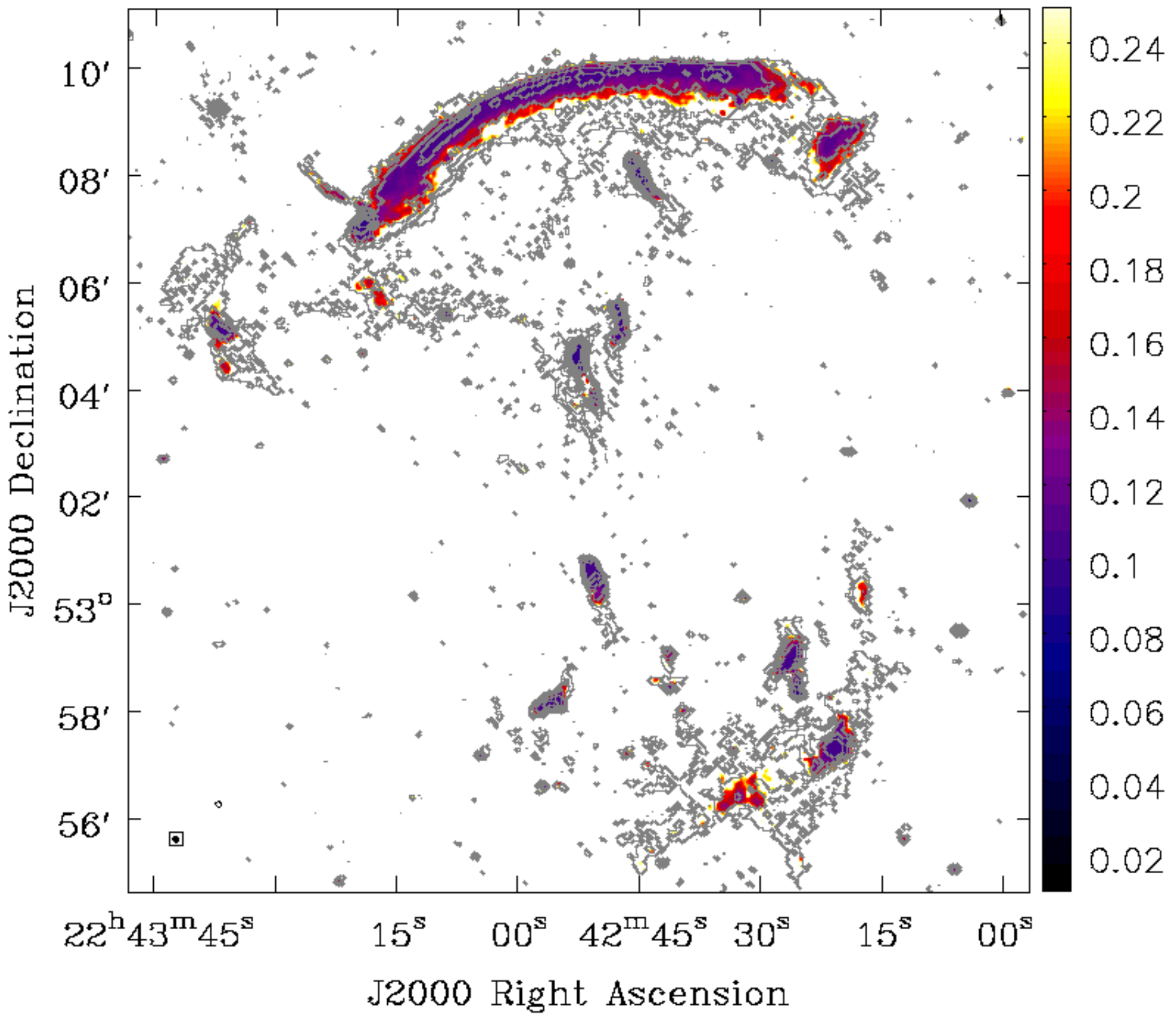}\caption{The corresponding spectral index error map for Fig. \ref{fig:SPX_HLres}.
\label{fig:SPX_Error_Map_Hres}}
\end{figure}
\par\end{center}

\begin{center}

\begin{figure}
\centering{}\includegraphics[width=1\columnwidth]{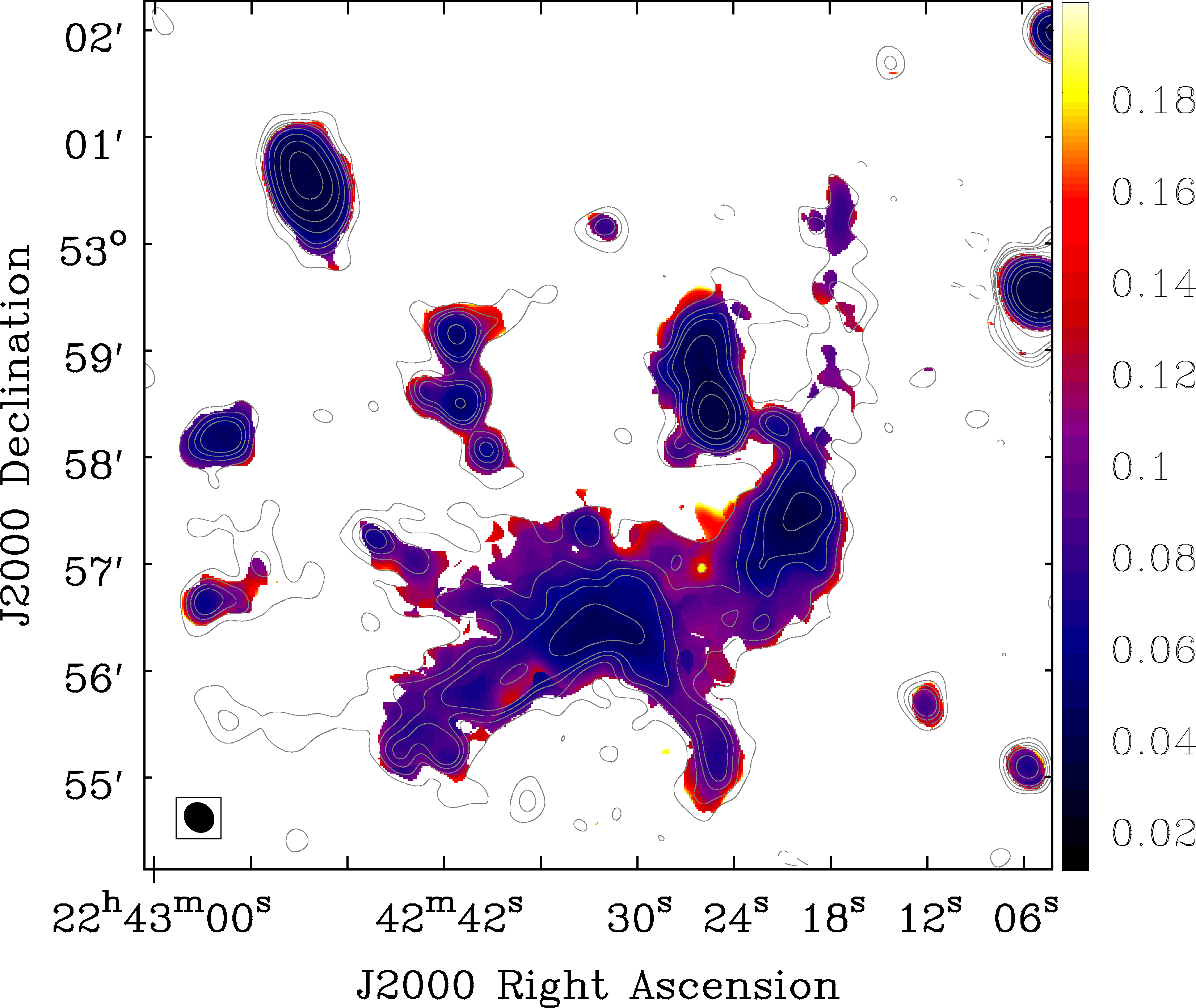}\caption{The corresponding spectral index error map for Fig. \ref{fig:SPX_RS_R1_R2}
(left). \label{fig:SPX_Error_Map_RS}}
\end{figure}

\begin{figure}
\centering{}\includegraphics[width=1\columnwidth]{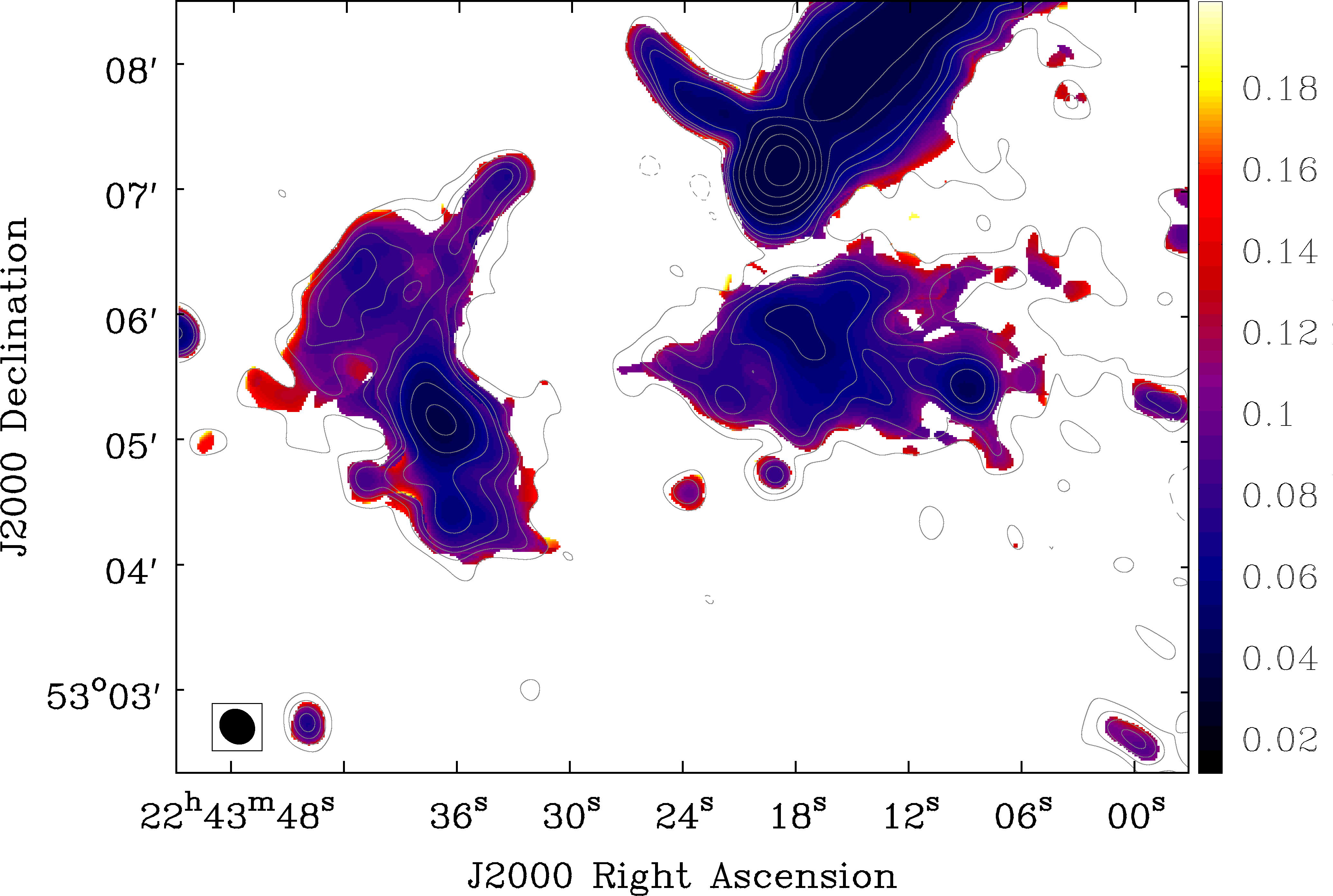}\caption{The corresponding spectral index error map for Fig. \ref{fig:SPX_RS_R1_R2}
(right). \label{fig:SPX_Error_Map_R1R2}}
\end{figure}

\begin{figure}
	\centering{}\includegraphics[width=1\columnwidth]{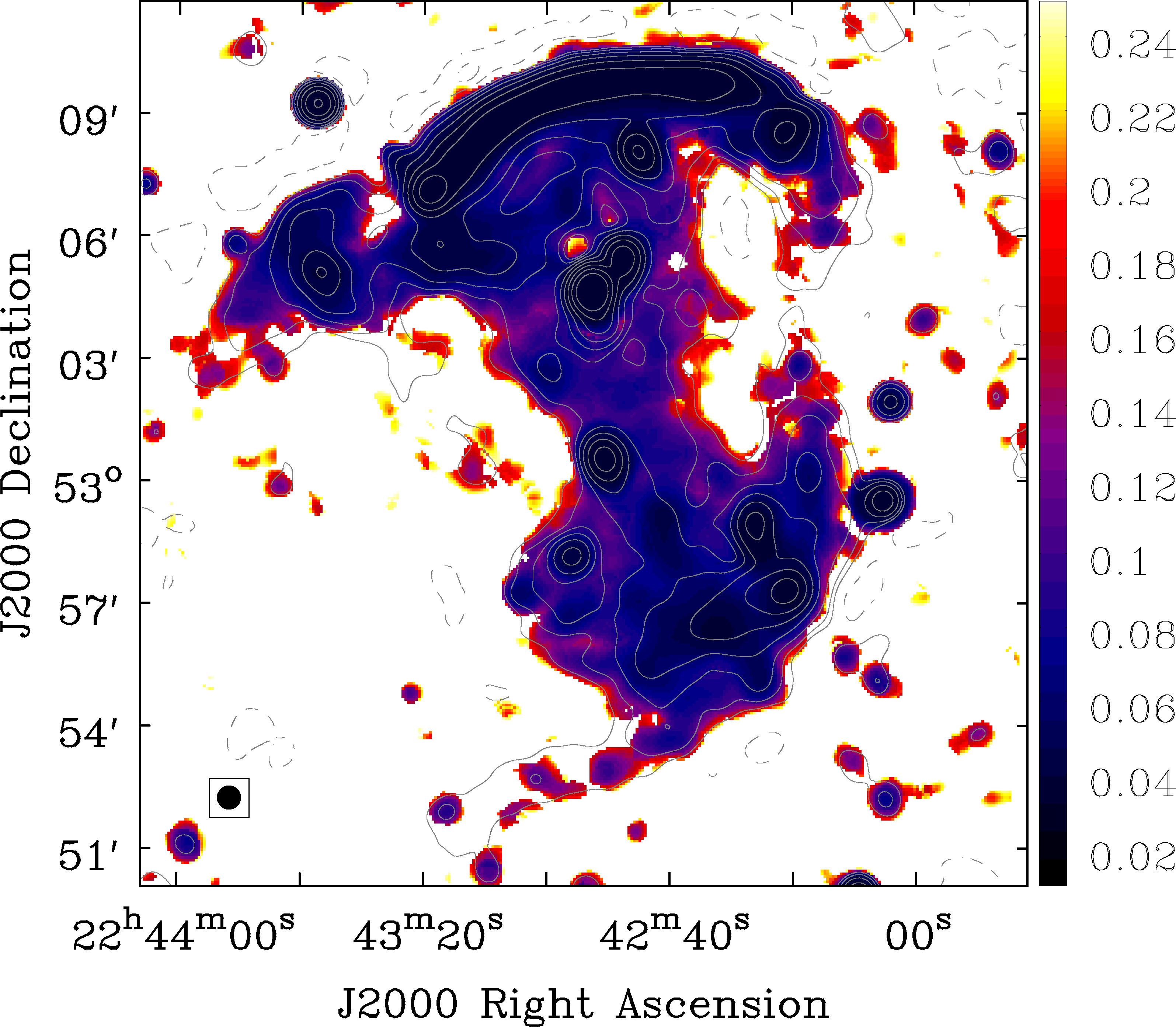}\caption{The corresponding spectral index error map for Fig. \ref{fig:Halo_Spx_profile}
		(left). \label{fig:SPX_Error_Map_Lres} }
\end{figure}

\par\end{center}

\section{Eastern region of RS}

A zoom-in optical image of the eastern region of RS is presented in
Fig. \ref{fig:SS_lowres_optical_radio}.
\begin{center}
\begin{figure}
\centering{}\includegraphics[width=1\columnwidth]{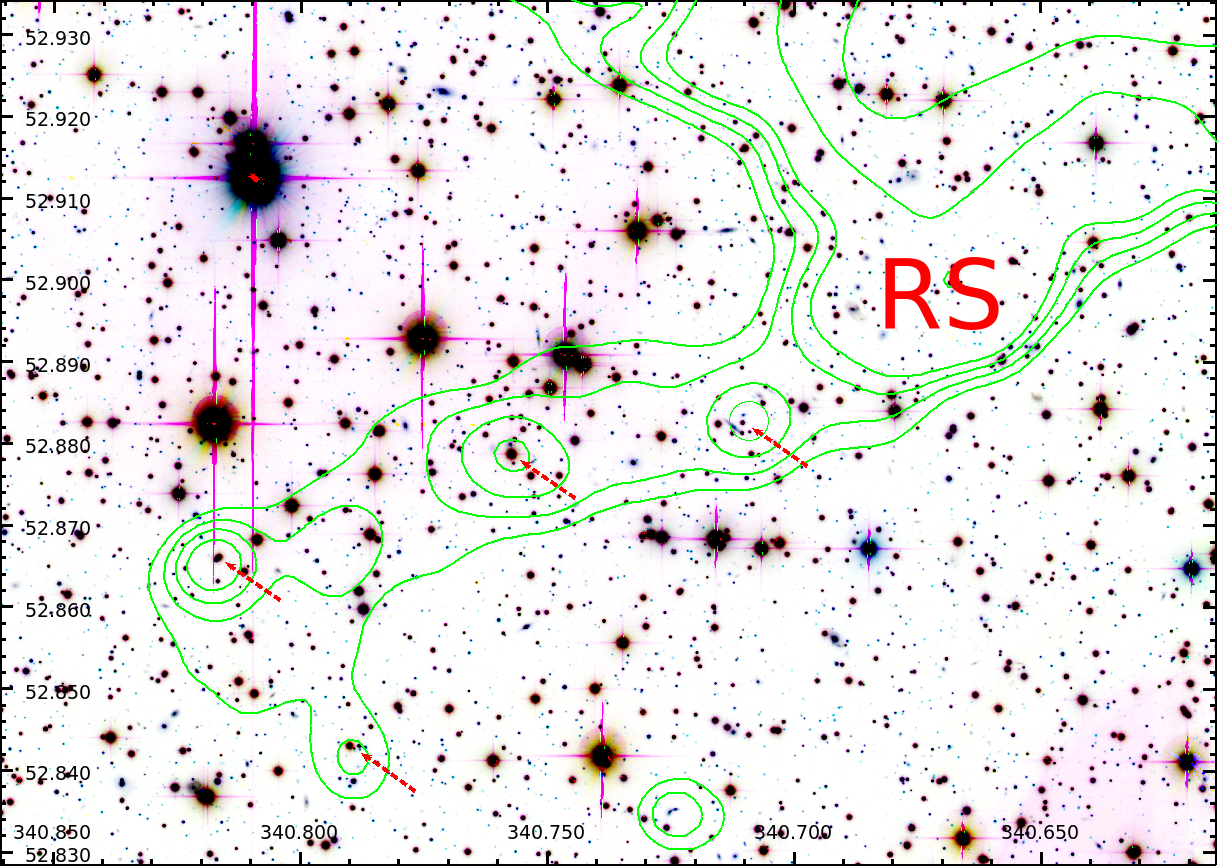}\caption{Optical image (Subaru \textit{g} and \textit{i} bands and CFHT\textit{
r} band, \citealt{Stroe2015h,Jee2015}) in the eastern side of RS.
Four optical counterparts (red arrows) are detected at the locations
of the radio emission peaks. The $35\arcsec$-resolution radio emission
contours are levelled at $[3,\,4,\,5,\,6,\,12,\,24]\times\sigma_{\text{noise}}$.
\label{fig:SS_lowres_optical_radio} }
\end{figure}
\par\end{center}








%
%


\bsp	
\label{lastpage}
\end{document}